\documentstyle[12pt]{article}

\begin{document}

\title{$\tau$ mesonic decays and strong anomaly of PCAC}
\author{Bing An Li\\
Department of Physics and Astronomy, University of Kentucky\\
Lexington, KY 40506, USA}

\maketitle

\begin{abstract}
It is shown that the matrix elements of the quark
axial-vector currents of
the decays $\tau\rightarrow\omega\pi\pi\nu$ and $\omega\rho\nu$ are not
conserved in the limit of \(m_{q}=0\). Pion exchange
dominates the decay mode $\tau\rightarrow\omega\rho\nu$. Theoretical
result of $\tau\rightarrow\omega(\pi\pi)_{non\rho}\nu$ agrees with
data well. Both the decay modes
provide evidences for the existence of strong anomaly of the PCAC.
The strong anomaly originates in the Wess-Zumino-Witten anomaly.
The PCAC with strong anomaly is written down and is applied to study
$\omega\rightarrow\pi\gamma$, $\rho\rightarrow\pi\gamma$, and $\omega
\rightarrow3\pi$ under the soft pion approximation. Theoretical results
are in good agreement with data. The decay $\tau\rightarrow K^{*}\rho
\nu$ and $K^{*}\omega\nu$ and the PCAC (\(\Delta s=1\)) with strong
anomaly is presented.

\end{abstract}

\newpage
The hadrons in $\tau$ hadronic decays are mesons, therefore
the $\tau$ mesonic decays provide a test ground for all meson
theories, especially the anomaly in meson physics.
$\pi^{0}\rightarrow2\gamma$ is via the PCAC related to
the Adler-Bell-Jackiw triangle anomaly[1].
The Adler-Bardeen theorem[2] is about the anomaly in $QCD$.
The Wess-Zumino-Witten(WZW)[3,4] Lagrangian
provides a general formalism for
various abnormal meson processes.
In Ref.[5] an effective chiral theory of mesons has been proposed.
In this theory the WZW Lagrangian is the leading term of the
imaginary part of the Lagrangian of the effective chiral theory.
The fields in the WZW Lagrangian are normalized to the physical meson
fields.
Based on chiral symmetry and chiral symmetry breaking,
a bosonized axial-vector currents of ordinary quarks has been presented
in our recent paper[6]. In this paper the vector currents of ordinary
quarks are treated by the VMD and
all the meson vertices are obtained
from the effective chiral theory[5]. The $\tau$ mesonic decays are
studied in terms of this theory[6]. Some of the decay modes are related
to the WZW anomaly. Theoretical results are in
reasonable agreement with data.
In the $\tau$-decay modes studied in [6] the $a_{1}$ meson is dominant
in the matrix elements of the axial-vector currents(u and d quarks)
and the axial-vector
currents are conserved in the chiral limit.

In this paper the decays $\tau\rightarrow\omega(\pi\pi)_{non\rho}\nu$
and $\tau
\rightarrow\omega\rho\nu$ are studied. Only axial-vector currents
contribute to these decays. It is found that the matrix elements of
the axial-vector currents are not conserved in the chiral limit and
pion dominates the decay $\tau\rightarrow\omega\rho\nu$. These
abnormal phenomena originate in the anomaly of PCAC. Here the anomaly
is the one of strong interaction. The strong interaction anomaly
of PCAC is studied in this paper.
The study on the decays $\tau\rightarrow K^{*}\rho\nu$ and $K^{*}
\omega\nu$ is presented. It is found in these two decays that
the axial-vector currents(
\(\Delta s=1\)) is not conserved in the limit of \(m_{q}=0\). The
strong anomaly of the PCAC(\(\Delta s=1 \)) is investigated.

In the chiral limit, the vector part of weak
interaction of ordinary quarks(u and d quarks only)
is written as[6]
\begin{equation}
{\cal L}^{V}={g_{W}\over 4}cos\theta_{C}
{1\over f_{\rho}}\{-{1\over 2}
(\partial_{\mu}A^{i}_{\nu}-\partial_{\nu}
A^{i}_{\mu})
(\partial_{\mu}\rho^{i}_
{\nu}-\partial_{\nu}\rho^{i}_{\mu})+A^{i}_{\mu}j^{i\mu}\},
\end{equation}
where \(i=1,2\) and $A^{i}_{\mu}$ are W boson fields,
$j^{i}_{\mu}$ is derived by using the substitution
\begin{equation}
\rho^{i}_{\mu}\rightarrow {g_{W}\over4f_{\rho}}cos\theta_{C}
A^{i}_{\mu}
\end{equation}
in the vertices involving $\rho$ mesons. In terms of the chiral symmetry
and spontaneous chiral symmetry breaking, the axial-vector part has
been determined as
\begin{eqnarray}
\lefteqn{{\cal L}^{A}=
-{g_{W}\over 4}cos\theta_{C}
{1\over f_{a}}\{-{1\over 2}(\partial_{\mu}A^{i}_{\nu}
-\partial_{\nu}A^{i}_{\mu})(\partial_{\mu}a^{i}_
{\nu}-\partial_{\nu}a^{i}_{\mu})+A^{i\mu}j^{iW}_{\mu}\}}\nonumber \\
&&-{g_{W}\over 4}cos\theta_{C}
\Delta m^{2}f_{a}A^{i}_{\mu}a^{i\mu}
-{g_{W}\over4}cos\theta_{C}
f_{\pi}A^{i}_{\mu}\partial^{\mu}\pi^{i},
\end{eqnarray}
where $f_{a}$ and $\Delta m^{2}$ are determined to be
\begin{equation}
f^{2}_{a}=f^{2}_{\rho}(1-{f^{2}_{\pi}f^{2}_{\rho}\over m^{2}_{\rho}})
{m^{2}_{a}\over m^{2}_{\rho}},\;\;\;
\Delta m^{2}=f^{2}_{\pi}(1-{f^{2}_{\pi}f^{2}_{\rho}\over
m^{2}_{\rho}})^{-1},
\end{equation}
$j^{i,W}_{\mu}$ is obtained by substituting
\begin{equation}
a^{i}_{\mu}\rightarrow-\frac{g_{W}}{4f_{a}}cos\theta_{C}A^{i}_{\mu}
\end{equation}
into the Lagrangian in which $a_{1}$ meson is involved.

The Lagrangians ${\cal L}^{V,A}$(1,3) have been derived from the
effective chiral theory of pseudoscalar, vector, and axial-vector
mesons[5].
The Lagrangian of this theory is expressed as
\begin{eqnarray}
{\cal L}=\bar{\psi}(x)(i\gamma\cdot\partial+\gamma\cdot v
+\gamma\cdot a\gamma_{5}
-mu(x))\psi(x)-\bar{\psi}M\psi\nonumber \\
+{1\over 2}m^{2}_{0}(\rho^{\mu}_{i}\rho_{\mu i}+
\omega^{\mu}\omega_{\mu}+a^{\mu}_{i}a_{\mu i}+f^{\mu}f_{\mu})
\end{eqnarray}
where \(a_{\mu}=\tau_{i}a^{i}_{\mu}+f_{\mu}\), \(v_{\mu}=\tau_{i}
\rho^{i}_{\mu}+\omega_{\mu}\),
, and \(u=exp\{i\gamma_{5}(\tau_{i}\pi_{i}+
\eta)\}\), these fields are normalized to
physical meson fields in Ref.[5].
The parameters of Eqs.(1,3) are defined as[5]
\begin{eqnarray}
\lefteqn{f_{\rho}=g^{-1},}\\
&&f_{a}=g^{-1}(1-{1\over2\pi^{2}g^{2}})^{-{1\over2}},\\
&&(1-{1\over 2\pi^{2}g^{2}})m^{2}_{a}=6m^{2}+m^{2}_{\rho},\\
&&\Delta m^{2}=6m^{2}g^{2}=
f^{2}_{\pi}(1-{f^{2}_{\pi}\over g^{2}m^{2}_{\rho}})^{-1},
\end{eqnarray}
where g is an universal coupling constant, \(g=0.39\)[6],
and m is a parameter related
to quark condensate[5].

It has been shown in Ref.[6] that
there are cancellations between the terms of Eq.(3) in the decay modes
studied and
these cancellations lead to both the conservation of the
axial-vector currents in the chiral limit
and the $a_{1}$ dominance.

The meson vertices involved in $\tau$ mesonic decays are obtained
from the effective Lagrangian of mesons presented in Ref.[5].
There are two kinds of vertices: the ones of normal parity and
the ones of abnormal parity. The later are the WZW anomaly.
Therefore, the theory of $\tau$ mesonic decays is completely
determined by the effective chiral theory of mesons[5].

The $\omega$ meson is
contained in the final states of both the decay modes
$\tau\rightarrow\omega\pi\pi\nu$ and $\tau\rightarrow\omega
\rho\nu$.
It is well known[4,5,7] that in two flavor case
if a vertex contains $\omega$-field,
the vertex is from WZW anomaly.
Therefore, both decay modes are related to the WZW anomaly.

The decay mode $\tau\rightarrow\omega\pi\pi\nu$ is composed of two
parts: the two pions are from a $\rho$ decay and the two pions are not
from a $\rho$ resonance. The study(see below) shows that the decay
rate of $\tau\rightarrow\omega(\pi\pi)_{\rho res.}\nu$ is smaller than
that of $\tau\rightarrow\omega(\pi\pi)_{non\rho}\nu$ by two order
of magnitude. We study $\tau\rightarrow\omega(\pi\pi)_{non\rho}\nu$
first.

Only the axial-vector currents(${\cal L}^{A}$(3)) contribute to
$\tau\rightarrow\omega(\pi\pi)_{non\rho}\nu$.
There are two kinds of vertices
involved in this decay channel. These vertices are derived from
the effective chiral theory of mesons[5] in the chiral limit.
\begin{enumerate}
\item The vertices $\omega\pi\pi\pi$ and $\omega a_{1}\pi\pi$
derived from Ref.[5] are
\begin{eqnarray}
\lefteqn{{\cal L}^{\omega\pi\pi\pi}
=\frac{2}{\pi^{2}gf^{3}_{\pi}}(1-{6c\over g}+{6c^{2}
\over
g^{2}})\varepsilon^{\mu\nu\alpha\beta}\epsilon_{ijk}\omega_{\mu}
\partial_{\nu}\pi_{i}\partial_{\alpha}\pi_{j}\partial_{\beta}\pi_{k}
}\nonumber \\
&&{\cal L}^{\omega a_{1}\pi\pi}=
-\frac{6}{\pi^{2}g^{2}f^{2}_{\pi}}(1-{1\over2\pi^{2}g^{2}})
^{-{1\over2}}(1-{2c\over g})\varepsilon^{\mu\nu\alpha\beta}
\epsilon_{ijk}\partial_{\nu}\omega_{\mu}a^{i}_{\alpha}
\partial_{\beta}\pi_{k},
\end{eqnarray}
where
\begin{equation}
c=\frac{f^{2}_{\pi}}{2gm^{2}_{\rho}}.
\end{equation}
These two vertices are from the WZW anomaly[5].
The vertex ${\cal L}^{W\omega\pi\pi}$ is found by using the
substitution(5) in the vertex ${\cal L}^{\omega a_{1}\pi\pi}$.
Using these vertices and ${\cal L}^{A}$(3),
the matrix element of the axial-vector current is obtained
\begin{eqnarray}
\lefteqn{<\omega\pi^{0}\pi^{-}|\bar{\psi}\tau_{+}\gamma_{\mu}
\gamma_{5}\psi|0>^{(1)}
=\frac{i}{\sqrt{8\omega_{1}\omega_{2}E}}\frac{6}
{\pi^{2}gf^{2}_{\pi}}\varepsilon^{\nu\lambda\alpha\beta}\epsilon^{*}_
{\lambda}p_{\alpha}(k_{2}-k_{1})_{\beta}}\nonumber \\
&&\{(\frac{q_{\mu}q_{\nu}}{q^{2}}-g_{\mu\nu})(1-{2c\over g})
\frac{g^{2}f^{2}_{a}m^{2}_{\rho}-i\sqrt{q^{2}}\Gamma_{a}(q^{2})}{
q^{2}-m^{2}_{a}+i\sqrt{q^{2}}\Gamma_{a}(q^{2})}
+\frac{q_{\mu}q_{\nu}}{q^{2}}{2c\over g}
(1-{2c\over g})\},
\end{eqnarray}
where $k_{1}$, $k_{2}$, and p are momentum of $\pi^{0}$, $\pi^{-}$, and
$\omega$ respectively, \(q=k_{1}+k_{2}+p\).
Eq.(13) shows that the matrix element obtained from the WZW anomaly is not
conserved in the limit of \(m_{q}=0\). The pion exchange is dominant
in the term which violates the conservation of the quark axial-vector
current in the chiral limit.
The divergence of this term
is written as
\begin{equation}
\frac{6}{\pi^{2}gf^{2}_{\pi}}{2c\over g}(1-{2c\over g})\varepsilon
^{\mu\nu\alpha\beta}\epsilon_{ijk}\partial_{\nu}\omega_{\mu}\partial
_{\alpha}\pi^{j}\partial_{\beta}\pi^{k}.
\end{equation}
\item The second kind of vertices are
vertices $a_{1}\rho\pi$, $\rho\pi\pi$, $W\rho\pi$
and $\omega\rho\pi$. The vertex $W\rho\pi$ is derived by substituting(5)
into the vertex $a_{1}\rho\pi$. The first three vertices have been
exploited to study $\tau\rightarrow\pi\pi\pi\nu$ and theoretical results
are in reasonably agreement with data.
The vertex $\omega\rho\pi$ is from the WZW anomaly.
\begin{eqnarray}
\lefteqn{{\cal L}^{a_{1}\rho\pi}=\epsilon_{ijk}\{Aa^{i}_{\mu}
\rho^{j\mu}\pi^{k}-Ba^{i}_{\mu}\rho^{j}_{\nu}\partial^{\mu\nu}\pi^{k}
+Da^{i}_{\mu}\partial^{\mu}(\rho^{j}_{\nu}
\partial^{\nu}\pi^{k})\}},\\
&&A={2\over f_{\pi}}gf_{a}\{
g^{2}f^{2}_{a}m^{2}_{a}-m^{2}_{\rho}+p^{2}[{2c\over g}+{3\over4
\pi^{2}g^{2}}(1-{2c\over g})]\nonumber \\
&&+q^{2}[{1\over 2\pi^{2}g^{2}}-
{2c\over g}-{3\over4\pi^{2}g^{2}}(1-{2c\over g})]\},\\
&&B=-{2\over f_{\pi}}gf_{a}{1\over2\pi^{2}g^{2}}(1-{2c\over g}),\\
&&D=-{2\over f_{\pi}}f_{a}\{2c+{3\over2\pi^{2}g}(1-{2c\over g})\},\\
&&{\cal L}^{\rho\pi\pi}={2\over g}\epsilon_{ijk}\rho^{i}_{\mu}
\pi^{j}\partial^{\mu}\pi^{k}-{2\over \pi^{2}f^{2}_{\pi}g}
\{(1-{2c\over g})^{2}-4\pi^{2}c^{2}\}\epsilon
_{ijk}\rho^{i}_{\mu}\partial_{\nu}\pi^{j}\partial^{\mu\nu}\pi^{k}
\nonumber \\
&&-{1\over \pi^{2}f^{2}_{\pi}g}\{3(1-{2c\over g})^{2}
+1-{2c\over g}-8\pi^{2}c^{2}\}\epsilon_{ijk}\rho^{i}_{\mu}\pi_{j}
\partial^{2}\partial_{\mu}\pi_{k}, \\
&&{\cal L}^{\omega\rho\pi}=
-\frac{N_{C}}{\pi^{2}g^{2}f_{\pi}}\varepsilon^{\mu\nu\alpha\beta}
\partial_{\mu}\omega_{\nu}\rho^{i}_{\alpha}\partial_{\beta}\pi^{i},
\end{eqnarray}
where p is the momentum of $\rho$ meson and q is
the momentum of $a_{1}$.
\end{enumerate}
Using the vertex ${\cal L}^{a_{1}\rho\pi}$(15), the decay width of
$a_{1}$ meson is derived
\begin{equation}
\Gamma_{a}(q^{2})
=\frac{k}{12\pi m_{a}\sqrt{q^{2}}}\{(3+{k^{2}\over
m^{2}_{\rho}})A^{2}-\frac{k^{2}}{m^{2}_{\rho}}
(q^{2}+m^{2}_{\rho})AB
+\frac{q^{2}}{m^{2}_{\rho}}k^{4}
B^{2}\},
\end{equation}
where
\[k=\{{1\over4q^{2}}(q^{2}+m^{2}_{\rho}-m^{2}_{\pi})^{2}-m^{2}_{\rho}\}
^{{1\over2}}.\]
Using ${\cal L}^{A}$(3)
and the vertices(15,19,20) and taking the cancellation
shown in Eq.(34) of Ref.[6] into account,
the second part of the matrix element of the
axial-vector current is obtained
\begin{eqnarray}
\lefteqn{<\omega\pi^{0}\pi^{-}|\bar{\psi}\tau_{+}\gamma_{\mu}\gamma_{5}
\psi|0>^{(2)}
=\frac{i}{\sqrt{8\omega_{1}\omega_{2}E}}(\frac{q_{\mu}q_{\nu}}
{q^{2}}-g_{\mu\nu})\frac{3}{\pi^{2}g^{2}
f_{\pi}}\frac{g^{2}f_{a}m^{2}
_{\rho}-if^{-1}_{a}
q\Gamma_{a}(q^{2})}{q^{2}-m^{2}_{a}+iq\Gamma_{a}(q^{2})}}
\nonumber \\
&&\varepsilon^{\nu\lambda\alpha\beta}\epsilon^{*}_{\lambda}p_{\alpha}
\{
\frac{A(k^{2})k_{2\beta}}{k^{2}-m^{2}_{\rho}+i\sqrt{k^{2}}
\Gamma_{\rho}(k^{2})}-\frac{A(k^{'2})k_{1\beta}}{k^{'2}-m^{2}_{\rho}+i
\sqrt{k^{'2}}\Gamma_{\rho}(k^{'2})}\}\nonumber \\
&&+\frac{i}{\sqrt{8\omega_{1}\omega_{2}E}}
(\frac{q_{\mu}q_{\nu}}
{q^{2}}-g_{\mu\nu})\frac{3}{\pi^{2}g^{2}
f_{\pi}}\frac{g^{2}f_{a}m^{2}
_{\rho}-iqf^{-1}_{a}\Gamma_{a}(q^{2})
}{q^{2}-m^{2}_{a}+iq\Gamma_{a}(q^{2})}
\varepsilon^{\sigma\lambda\alpha\beta}
\epsilon^{*}_{\sigma}p_{\lambda}k_{2\alpha}k_{1\beta}(-B)\nonumber \\
&&\{\frac{k_{2\nu}}{k^{'2}-m^{2}_{\rho}+i\sqrt{k^{'2}}\Gamma_{\rho}
(k^{'2})}+\frac{k_{1\nu}}{k^{2}-m^{2}_{\rho}+i\sqrt{k^{2}}\Gamma_{\rho}
(k^{2})}\},
\end{eqnarray}
where $k_{1}$, $k_{2}$, p are momentum of $\pi^{0}$, $\pi^{-}$, and
$\omega$ mesons respectively,
\(q=p+k_{1}+k_{2}\), \(k=q-k_{1}\), \(k^{'}=q-k_{2}\),
$A(k^{2})$ and $A(k^{'2})$ are defined by Eq.(16) by taking
\(p^{2}=k^{2}, k^{'2}\) respectively.
This part of the matrix element observes the axial-vector current
conservation in the limit \(m_{q}=0\) and there is $a_{1}$ dominance.

Adding Eqs.(13,22) together, the whole matrix element is obtained.
The expression of the decay
width is presented in the Appendix.
The branching ratio is computed to be
\[B=0.37\%.\]
The data is $0.41\pm0.08\pm0.06\%$[7]. The distribution of the decay
rate versus the invariant mass of $\omega\pi\pi$ is shown in Fig.1.

\begin{figure}
\begin{center}

\setlength{\unitlength}{0.240900pt}
\ifx\plotpoint\undefined\newsavebox{\plotpoint}\fi
\sbox{\plotpoint}{\rule[-0.500pt]{1.000pt}{1.000pt}}%
\begin{picture}(1500,900)(0,0)
\font\gnuplot=cmr10 at 10pt
\gnuplot
\sbox{\plotpoint}{\rule[-0.500pt]{1.000pt}{1.000pt}}%
\put(220.0,113.0){\rule[-0.500pt]{292.934pt}{1.000pt}}
\put(220.0,113.0){\rule[-0.500pt]{4.818pt}{1.000pt}}
\put(198,113){\makebox(0,0)[r]{0}}
\put(1416.0,113.0){\rule[-0.500pt]{4.818pt}{1.000pt}}
\put(220.0,222.0){\rule[-0.500pt]{4.818pt}{1.000pt}}
\put(198,222){\makebox(0,0)[r]{10}}
\put(1416.0,222.0){\rule[-0.500pt]{4.818pt}{1.000pt}}
\put(220.0,331.0){\rule[-0.500pt]{4.818pt}{1.000pt}}
\put(198,331){\makebox(0,0)[r]{20}}
\put(1416.0,331.0){\rule[-0.500pt]{4.818pt}{1.000pt}}
\put(220.0,440.0){\rule[-0.500pt]{4.818pt}{1.000pt}}
\put(198,440){\makebox(0,0)[r]{30}}
\put(1416.0,440.0){\rule[-0.500pt]{4.818pt}{1.000pt}}
\put(220.0,550.0){\rule[-0.500pt]{4.818pt}{1.000pt}}
\put(198,550){\makebox(0,0)[r]{40}}
\put(1416.0,550.0){\rule[-0.500pt]{4.818pt}{1.000pt}}
\put(220.0,659.0){\rule[-0.500pt]{4.818pt}{1.000pt}}
\put(198,659){\makebox(0,0)[r]{50}}
\put(1416.0,659.0){\rule[-0.500pt]{4.818pt}{1.000pt}}
\put(220.0,768.0){\rule[-0.500pt]{4.818pt}{1.000pt}}
\put(198,768){\makebox(0,0)[r]{60}}
\put(1416.0,768.0){\rule[-0.500pt]{4.818pt}{1.000pt}}
\put(220.0,877.0){\rule[-0.500pt]{4.818pt}{1.000pt}}
\put(198,877){\makebox(0,0)[r]{70}}
\put(1416.0,877.0){\rule[-0.500pt]{4.818pt}{1.000pt}}
\put(220.0,113.0){\rule[-0.500pt]{1.000pt}{4.818pt}}
\put(220,68){\makebox(0,0){0.8}}
\put(220.0,857.0){\rule[-0.500pt]{1.000pt}{4.818pt}}
\put(423.0,113.0){\rule[-0.500pt]{1.000pt}{4.818pt}}
\put(423,68){\makebox(0,0){1}}
\put(423.0,857.0){\rule[-0.500pt]{1.000pt}{4.818pt}}
\put(625.0,113.0){\rule[-0.500pt]{1.000pt}{4.818pt}}
\put(625,68){\makebox(0,0){1.2}}
\put(625.0,857.0){\rule[-0.500pt]{1.000pt}{4.818pt}}
\put(828.0,113.0){\rule[-0.500pt]{1.000pt}{4.818pt}}
\put(828,68){\makebox(0,0){1.4}}
\put(828.0,857.0){\rule[-0.500pt]{1.000pt}{4.818pt}}
\put(1031.0,113.0){\rule[-0.500pt]{1.000pt}{4.818pt}}
\put(1031,68){\makebox(0,0){1.6}}
\put(1031.0,857.0){\rule[-0.500pt]{1.000pt}{4.818pt}}
\put(1233.0,113.0){\rule[-0.500pt]{1.000pt}{4.818pt}}
\put(1233,68){\makebox(0,0){1.8}}
\put(1233.0,857.0){\rule[-0.500pt]{1.000pt}{4.818pt}}
\put(1436.0,113.0){\rule[-0.500pt]{1.000pt}{4.818pt}}
\put(1436,68){\makebox(0,0){2}}
\put(1436.0,857.0){\rule[-0.500pt]{1.000pt}{4.818pt}}
\put(220.0,113.0){\rule[-0.500pt]{292.934pt}{1.000pt}}
\put(1436.0,113.0){\rule[-0.500pt]{1.000pt}{184.048pt}}
\put(220.0,877.0){\rule[-0.500pt]{292.934pt}{1.000pt}}
\put(45,495){\makebox(0,0){${d\Gamma\over d\sqrt{q^{2}}}\times 10^{15}$ }}
\put(828,23){\makebox(0,0){FiG.1          $\sqrt{q^{2}}$      GeV}}
\put(220.0,113.0){\rule[-0.500pt]{1.000pt}{184.048pt}}
\put(502,122){\usebox{\plotpoint}}
\multiput(502.00,123.84)(0.579,0.462){4}{\rule{1.750pt}{0.111pt}}
\multiput(502.00,119.92)(5.368,6.000){2}{\rule{0.875pt}{1.000pt}}
\multiput(511.00,129.84)(0.606,0.475){6}{\rule{1.679pt}{0.114pt}}
\multiput(511.00,125.92)(6.516,7.000){2}{\rule{0.839pt}{1.000pt}}
\multiput(521.00,136.83)(0.423,0.485){10}{\rule{1.250pt}{0.117pt}}
\multiput(521.00,132.92)(6.406,9.000){2}{\rule{0.625pt}{1.000pt}}
\multiput(531.83,144.00)(0.485,0.483){10}{\rule{0.117pt}{1.361pt}}
\multiput(527.92,144.00)(9.000,7.175){2}{\rule{1.000pt}{0.681pt}}
\multiput(540.83,154.00)(0.485,0.603){10}{\rule{0.117pt}{1.583pt}}
\multiput(536.92,154.00)(9.000,8.714){2}{\rule{1.000pt}{0.792pt}}
\multiput(549.83,166.00)(0.485,0.783){10}{\rule{0.117pt}{1.917pt}}
\multiput(545.92,166.00)(9.000,11.022){2}{\rule{1.000pt}{0.958pt}}
\multiput(558.83,181.00)(0.485,0.842){10}{\rule{0.117pt}{2.028pt}}
\multiput(554.92,181.00)(9.000,11.791){2}{\rule{1.000pt}{1.014pt}}
\multiput(567.83,197.00)(0.485,1.082){10}{\rule{0.117pt}{2.472pt}}
\multiput(563.92,197.00)(9.000,14.869){2}{\rule{1.000pt}{1.236pt}}
\multiput(576.83,217.00)(0.485,1.202){10}{\rule{0.117pt}{2.694pt}}
\multiput(572.92,217.00)(9.000,16.408){2}{\rule{1.000pt}{1.347pt}}
\multiput(585.83,239.00)(0.481,1.639){8}{\rule{0.116pt}{3.500pt}}
\multiput(581.92,239.00)(8.000,18.736){2}{\rule{1.000pt}{1.750pt}}
\multiput(593.83,265.00)(0.485,1.681){10}{\rule{0.117pt}{3.583pt}}
\multiput(589.92,265.00)(9.000,22.563){2}{\rule{1.000pt}{1.792pt}}
\multiput(602.83,295.00)(0.485,1.860){10}{\rule{0.117pt}{3.917pt}}
\multiput(598.92,295.00)(9.000,24.871){2}{\rule{1.000pt}{1.958pt}}
\multiput(611.83,328.00)(0.481,2.395){8}{\rule{0.116pt}{4.875pt}}
\multiput(607.92,328.00)(8.000,26.882){2}{\rule{1.000pt}{2.438pt}}
\multiput(619.83,365.00)(0.481,2.670){8}{\rule{0.116pt}{5.375pt}}
\multiput(615.92,365.00)(8.000,29.844){2}{\rule{1.000pt}{2.688pt}}
\multiput(627.83,406.00)(0.485,2.519){10}{\rule{0.117pt}{5.139pt}}
\multiput(623.92,406.00)(9.000,33.334){2}{\rule{1.000pt}{2.569pt}}
\multiput(636.83,450.00)(0.481,3.013){8}{\rule{0.116pt}{6.000pt}}
\multiput(632.92,450.00)(8.000,33.547){2}{\rule{1.000pt}{3.000pt}}
\multiput(644.83,496.00)(0.485,2.638){10}{\rule{0.117pt}{5.361pt}}
\multiput(640.92,496.00)(9.000,34.873){2}{\rule{1.000pt}{2.681pt}}
\multiput(653.83,542.00)(0.481,3.082){8}{\rule{0.116pt}{6.125pt}}
\multiput(649.92,542.00)(8.000,34.287){2}{\rule{1.000pt}{3.062pt}}
\multiput(661.83,589.00)(0.481,2.876){8}{\rule{0.116pt}{5.750pt}}
\multiput(657.92,589.00)(8.000,32.066){2}{\rule{1.000pt}{2.875pt}}
\multiput(669.83,633.00)(0.481,2.601){8}{\rule{0.116pt}{5.250pt}}
\multiput(665.92,633.00)(8.000,29.103){2}{\rule{1.000pt}{2.625pt}}
\multiput(677.83,673.00)(0.481,2.189){8}{\rule{0.116pt}{4.500pt}}
\multiput(673.92,673.00)(8.000,24.660){2}{\rule{1.000pt}{2.250pt}}
\multiput(685.83,707.00)(0.481,1.707){8}{\rule{0.116pt}{3.625pt}}
\multiput(681.92,707.00)(8.000,19.476){2}{\rule{1.000pt}{1.813pt}}
\multiput(693.83,734.00)(0.481,1.226){8}{\rule{0.116pt}{2.750pt}}
\multiput(689.92,734.00)(8.000,14.292){2}{\rule{1.000pt}{1.375pt}}
\multiput(701.83,754.00)(0.481,0.677){8}{\rule{0.116pt}{1.750pt}}
\multiput(697.92,754.00)(8.000,8.368){2}{\rule{1.000pt}{0.875pt}}
\put(708,765.92){\rule{1.927pt}{1.000pt}}
\multiput(708.00,763.92)(4.000,4.000){2}{\rule{0.964pt}{1.000pt}}
\put(716,766.42){\rule{1.927pt}{1.000pt}}
\multiput(716.00,767.92)(4.000,-3.000){2}{\rule{0.964pt}{1.000pt}}
\multiput(724.00,764.68)(0.402,-0.481){8}{\rule{1.250pt}{0.116pt}}
\multiput(724.00,764.92)(5.406,-8.000){2}{\rule{0.625pt}{1.000pt}}
\multiput(733.84,750.25)(0.475,-0.851){6}{\rule{0.114pt}{2.107pt}}
\multiput(729.92,754.63)(7.000,-8.627){2}{\rule{1.000pt}{1.054pt}}
\multiput(740.83,736.14)(0.481,-1.020){8}{\rule{0.116pt}{2.375pt}}
\multiput(736.92,741.07)(8.000,-12.071){2}{\rule{1.000pt}{1.188pt}}
\multiput(748.83,718.10)(0.481,-1.158){8}{\rule{0.116pt}{2.625pt}}
\multiput(744.92,723.55)(8.000,-13.552){2}{\rule{1.000pt}{1.313pt}}
\multiput(756.84,696.51)(0.475,-1.502){6}{\rule{0.114pt}{3.250pt}}
\multiput(752.92,703.25)(7.000,-14.254){2}{\rule{1.000pt}{1.625pt}}
\multiput(763.83,677.07)(0.481,-1.295){8}{\rule{0.116pt}{2.875pt}}
\multiput(759.92,683.03)(8.000,-15.033){2}{\rule{1.000pt}{1.438pt}}
\multiput(771.84,653.92)(0.475,-1.583){6}{\rule{0.114pt}{3.393pt}}
\multiput(767.92,660.96)(7.000,-14.958){2}{\rule{1.000pt}{1.696pt}}
\multiput(778.83,633.55)(0.481,-1.364){8}{\rule{0.116pt}{3.000pt}}
\multiput(774.92,639.77)(8.000,-15.773){2}{\rule{1.000pt}{1.500pt}}
\multiput(786.84,610.51)(0.475,-1.502){6}{\rule{0.114pt}{3.250pt}}
\multiput(782.92,617.25)(7.000,-14.254){2}{\rule{1.000pt}{1.625pt}}
\multiput(793.83,591.07)(0.481,-1.295){8}{\rule{0.116pt}{2.875pt}}
\multiput(789.92,597.03)(8.000,-15.033){2}{\rule{1.000pt}{1.438pt}}
\multiput(801.84,569.10)(0.475,-1.420){6}{\rule{0.114pt}{3.107pt}}
\multiput(797.92,575.55)(7.000,-13.551){2}{\rule{1.000pt}{1.554pt}}
\multiput(808.84,549.10)(0.475,-1.420){6}{\rule{0.114pt}{3.107pt}}
\multiput(804.92,555.55)(7.000,-13.551){2}{\rule{1.000pt}{1.554pt}}
\multiput(815.83,531.62)(0.481,-1.089){8}{\rule{0.116pt}{2.500pt}}
\multiput(811.92,536.81)(8.000,-12.811){2}{\rule{1.000pt}{1.250pt}}
\multiput(823.84,512.29)(0.475,-1.258){6}{\rule{0.114pt}{2.821pt}}
\multiput(819.92,518.14)(7.000,-12.144){2}{\rule{1.000pt}{1.411pt}}
\multiput(830.84,494.88)(0.475,-1.176){6}{\rule{0.114pt}{2.679pt}}
\multiput(826.92,500.44)(7.000,-11.440){2}{\rule{1.000pt}{1.339pt}}
\multiput(837.84,478.47)(0.475,-1.095){6}{\rule{0.114pt}{2.536pt}}
\multiput(833.92,483.74)(7.000,-10.737){2}{\rule{1.000pt}{1.268pt}}
\multiput(844.84,463.07)(0.475,-1.013){6}{\rule{0.114pt}{2.393pt}}
\multiput(840.92,468.03)(7.000,-10.034){2}{\rule{1.000pt}{1.196pt}}
\multiput(851.84,448.66)(0.475,-0.932){6}{\rule{0.114pt}{2.250pt}}
\multiput(847.92,453.33)(7.000,-9.330){2}{\rule{1.000pt}{1.125pt}}
\multiput(858.83,435.70)(0.481,-0.814){8}{\rule{0.116pt}{2.000pt}}
\multiput(854.92,439.85)(8.000,-9.849){2}{\rule{1.000pt}{1.000pt}}
\multiput(866.84,421.25)(0.475,-0.851){6}{\rule{0.114pt}{2.107pt}}
\multiput(862.92,425.63)(7.000,-8.627){2}{\rule{1.000pt}{1.054pt}}
\multiput(873.84,408.25)(0.475,-0.851){6}{\rule{0.114pt}{2.107pt}}
\multiput(869.92,412.63)(7.000,-8.627){2}{\rule{1.000pt}{1.054pt}}
\multiput(880.84,395.85)(0.475,-0.769){6}{\rule{0.114pt}{1.964pt}}
\multiput(876.92,399.92)(7.000,-7.923){2}{\rule{1.000pt}{0.982pt}}
\multiput(887.84,383.35)(0.462,-0.784){4}{\rule{0.111pt}{2.083pt}}
\multiput(883.92,387.68)(6.000,-6.676){2}{\rule{1.000pt}{1.042pt}}
\multiput(893.84,373.44)(0.475,-0.688){6}{\rule{0.114pt}{1.821pt}}
\multiput(889.92,377.22)(7.000,-7.220){2}{\rule{1.000pt}{0.911pt}}
\multiput(900.84,363.03)(0.475,-0.606){6}{\rule{0.114pt}{1.679pt}}
\multiput(896.92,366.52)(7.000,-6.516){2}{\rule{1.000pt}{0.839pt}}
\multiput(907.84,353.03)(0.475,-0.606){6}{\rule{0.114pt}{1.679pt}}
\multiput(903.92,356.52)(7.000,-6.516){2}{\rule{1.000pt}{0.839pt}}
\multiput(914.84,343.03)(0.475,-0.606){6}{\rule{0.114pt}{1.679pt}}
\multiput(910.92,346.52)(7.000,-6.516){2}{\rule{1.000pt}{0.839pt}}
\multiput(921.84,333.63)(0.475,-0.525){6}{\rule{0.114pt}{1.536pt}}
\multiput(917.92,336.81)(7.000,-5.813){2}{\rule{1.000pt}{0.768pt}}
\multiput(928.84,323.74)(0.462,-0.579){4}{\rule{0.111pt}{1.750pt}}
\multiput(924.92,327.37)(6.000,-5.368){2}{\rule{1.000pt}{0.875pt}}
\multiput(934.84,315.63)(0.475,-0.525){6}{\rule{0.114pt}{1.536pt}}
\multiput(930.92,318.81)(7.000,-5.813){2}{\rule{1.000pt}{0.768pt}}
\multiput(941.84,306.63)(0.475,-0.525){6}{\rule{0.114pt}{1.536pt}}
\multiput(937.92,309.81)(7.000,-5.813){2}{\rule{1.000pt}{0.768pt}}
\multiput(948.84,297.43)(0.462,-0.476){4}{\rule{0.111pt}{1.583pt}}
\multiput(944.92,300.71)(6.000,-4.714){2}{\rule{1.000pt}{0.792pt}}
\multiput(954.84,290.22)(0.475,-0.444){6}{\rule{0.114pt}{1.393pt}}
\multiput(950.92,293.11)(7.000,-5.109){2}{\rule{1.000pt}{0.696pt}}
\multiput(960.00,285.69)(0.362,-0.475){6}{\rule{1.250pt}{0.114pt}}
\multiput(960.00,285.92)(4.406,-7.000){2}{\rule{0.625pt}{1.000pt}}
\multiput(968.84,274.43)(0.462,-0.476){4}{\rule{0.111pt}{1.583pt}}
\multiput(964.92,277.71)(6.000,-4.714){2}{\rule{1.000pt}{0.792pt}}
\multiput(973.00,270.69)(0.362,-0.475){6}{\rule{1.250pt}{0.114pt}}
\multiput(973.00,270.92)(4.406,-7.000){2}{\rule{0.625pt}{1.000pt}}
\multiput(981.84,260.12)(0.462,-0.373){4}{\rule{0.111pt}{1.417pt}}
\multiput(977.92,263.06)(6.000,-4.060){2}{\rule{1.000pt}{0.708pt}}
\multiput(986.00,256.69)(0.362,-0.475){6}{\rule{1.250pt}{0.114pt}}
\multiput(986.00,256.92)(4.406,-7.000){2}{\rule{0.625pt}{1.000pt}}
\multiput(994.84,246.12)(0.462,-0.373){4}{\rule{0.111pt}{1.417pt}}
\multiput(990.92,249.06)(6.000,-4.060){2}{\rule{1.000pt}{0.708pt}}
\multiput(999.00,242.69)(0.362,-0.475){6}{\rule{1.250pt}{0.114pt}}
\multiput(999.00,242.92)(4.406,-7.000){2}{\rule{0.625pt}{1.000pt}}
\multiput(1006.00,235.69)(0.270,-0.462){4}{\rule{1.250pt}{0.111pt}}
\multiput(1006.00,235.92)(3.406,-6.000){2}{\rule{0.625pt}{1.000pt}}
\multiput(1012.00,229.69)(0.373,-0.462){4}{\rule{1.417pt}{0.111pt}}
\multiput(1012.00,229.92)(4.060,-6.000){2}{\rule{0.708pt}{1.000pt}}
\multiput(1020.84,220.12)(0.462,-0.373){4}{\rule{0.111pt}{1.417pt}}
\multiput(1016.92,223.06)(6.000,-4.060){2}{\rule{1.000pt}{0.708pt}}
\multiput(1025.00,216.69)(0.270,-0.462){4}{\rule{1.250pt}{0.111pt}}
\multiput(1025.00,216.92)(3.406,-6.000){2}{\rule{0.625pt}{1.000pt}}
\multiput(1031.00,210.71)(0.151,-0.424){2}{\rule{1.650pt}{0.102pt}}
\multiput(1031.00,210.92)(3.575,-5.000){2}{\rule{0.825pt}{1.000pt}}
\multiput(1038.00,205.69)(0.270,-0.462){4}{\rule{1.250pt}{0.111pt}}
\multiput(1038.00,205.92)(3.406,-6.000){2}{\rule{0.625pt}{1.000pt}}
\multiput(1044.00,199.69)(0.270,-0.462){4}{\rule{1.250pt}{0.111pt}}
\multiput(1044.00,199.92)(3.406,-6.000){2}{\rule{0.625pt}{1.000pt}}
\put(1050,191.42){\rule{1.445pt}{1.000pt}}
\multiput(1050.00,193.92)(3.000,-5.000){2}{\rule{0.723pt}{1.000pt}}
\multiput(1056.00,188.69)(0.373,-0.462){4}{\rule{1.417pt}{0.111pt}}
\multiput(1056.00,188.92)(4.060,-6.000){2}{\rule{0.708pt}{1.000pt}}
\put(1063,180.42){\rule{1.445pt}{1.000pt}}
\multiput(1063.00,182.92)(3.000,-5.000){2}{\rule{0.723pt}{1.000pt}}
\put(1069,175.42){\rule{1.445pt}{1.000pt}}
\multiput(1069.00,177.92)(3.000,-5.000){2}{\rule{0.723pt}{1.000pt}}
\put(1075,170.42){\rule{1.445pt}{1.000pt}}
\multiput(1075.00,172.92)(3.000,-5.000){2}{\rule{0.723pt}{1.000pt}}
\put(1081,165.92){\rule{1.445pt}{1.000pt}}
\multiput(1081.00,167.92)(3.000,-4.000){2}{\rule{0.723pt}{1.000pt}}
\put(1087,161.42){\rule{1.445pt}{1.000pt}}
\multiput(1087.00,163.92)(3.000,-5.000){2}{\rule{0.723pt}{1.000pt}}
\put(1093,156.92){\rule{1.445pt}{1.000pt}}
\multiput(1093.00,158.92)(3.000,-4.000){2}{\rule{0.723pt}{1.000pt}}
\put(1099,152.92){\rule{1.445pt}{1.000pt}}
\multiput(1099.00,154.92)(3.000,-4.000){2}{\rule{0.723pt}{1.000pt}}
\put(1105,148.92){\rule{1.445pt}{1.000pt}}
\multiput(1105.00,150.92)(3.000,-4.000){2}{\rule{0.723pt}{1.000pt}}
\put(1111,144.92){\rule{1.686pt}{1.000pt}}
\multiput(1111.00,146.92)(3.500,-4.000){2}{\rule{0.843pt}{1.000pt}}
\put(1118,140.92){\rule{1.204pt}{1.000pt}}
\multiput(1118.00,142.92)(2.500,-4.000){2}{\rule{0.602pt}{1.000pt}}
\put(1123,137.42){\rule{1.445pt}{1.000pt}}
\multiput(1123.00,138.92)(3.000,-3.000){2}{\rule{0.723pt}{1.000pt}}
\put(1129,133.92){\rule{1.445pt}{1.000pt}}
\multiput(1129.00,135.92)(3.000,-4.000){2}{\rule{0.723pt}{1.000pt}}
\put(1135,130.42){\rule{1.445pt}{1.000pt}}
\multiput(1135.00,131.92)(3.000,-3.000){2}{\rule{0.723pt}{1.000pt}}
\put(1141,127.92){\rule{1.445pt}{1.000pt}}
\multiput(1141.00,128.92)(3.000,-2.000){2}{\rule{0.723pt}{1.000pt}}
\put(1147,125.42){\rule{1.445pt}{1.000pt}}
\multiput(1147.00,126.92)(3.000,-3.000){2}{\rule{0.723pt}{1.000pt}}
\put(1153,122.92){\rule{1.445pt}{1.000pt}}
\multiput(1153.00,123.92)(3.000,-2.000){2}{\rule{0.723pt}{1.000pt}}
\put(1159,120.42){\rule{1.445pt}{1.000pt}}
\multiput(1159.00,121.92)(3.000,-3.000){2}{\rule{0.723pt}{1.000pt}}
\put(1165,117.92){\rule{1.445pt}{1.000pt}}
\multiput(1165.00,118.92)(3.000,-2.000){2}{\rule{0.723pt}{1.000pt}}
\put(1171,116.42){\rule{1.445pt}{1.000pt}}
\multiput(1171.00,116.92)(3.000,-1.000){2}{\rule{0.723pt}{1.000pt}}
\put(1177,114.92){\rule{1.204pt}{1.000pt}}
\multiput(1177.00,115.92)(2.500,-2.000){2}{\rule{0.602pt}{1.000pt}}
\put(1182,113.42){\rule{1.445pt}{1.000pt}}
\multiput(1182.00,113.92)(3.000,-1.000){2}{\rule{0.723pt}{1.000pt}}
\put(1188,112.42){\rule{1.445pt}{1.000pt}}
\multiput(1188.00,112.92)(3.000,-1.000){2}{\rule{0.723pt}{1.000pt}}
\put(1200,111.42){\rule{1.204pt}{1.000pt}}
\multiput(1200.00,111.92)(2.500,-1.000){2}{\rule{0.602pt}{1.000pt}}
\put(1194.0,114.0){\rule[-0.500pt]{1.445pt}{1.000pt}}
\end{picture}

\end{center}
\end{figure}

It is the same as the decay mode studied above,
only the axial-vector currents contribute to
$\tau\rightarrow\omega\rho\nu$.
At the tree level ${\cal L}^{\omega\rho\pi}$
(20)is the only vertex involved in this decay channel.
Therefore, the decay is resulted by the WZW anomaly.
Using the term
\[-{g_{W}\over 4}cos\theta_{C}f_{\pi}A^{i}_{\mu}
\partial^{\mu}\pi^{i}\]
in Eq.(3) and the vertex(20), the matrix element of the axial-vector
current is obtained
\begin{equation}
<\omega\rho^{-}|\bar{\psi}\tau_{+}\gamma_{\mu}\gamma_{5}\psi|0>=
-\frac{i}{\sqrt{4E_{1}E_{2}}}
\frac{N_{C}}{\pi^{2}g^{2}}\frac{q_{\mu}}{q^{2}}
\varepsilon^{\lambda\nu\alpha\beta}p_{1\lambda}p_{2\nu}\epsilon^{*}_{
\alpha}(p_{1})\epsilon^{*}_{\beta}(p_{2}).
\end{equation}
In the limit \(m_{q}=0\), the axial-vector current is not conserved
in this process.
The conservation of the
matrix element(22) is resulted by the cancellations between the terms
${1\over f_{a}}j^{iW}_{\mu}$, $\Delta m^{2}f_{a}a^{i}_{\mu}$, and
$f_{\pi}\partial_{\mu}\pi^{i}$ of ${\cal L}^{A}$(5)(see Eq.(34) of
Ref.[6]). In the matrix element(13) there is cancellation between
the two vertices(11), however, the cancellation is not enough.
This result leads to the nonconservation of the axial-vector
current in Eq.(13). For the process(23)
there is no contribution from $a_{1}$ meson,
therefore, there is no such cancellation. The pion exchange dominates
this decay and the axial-vector current is not conserved.
The divergence of the operator used to derive the matrix element(23)
is
\begin{equation}
\frac{N_{C}}{\pi^{2}g^{2}}\varepsilon^{\mu\nu\alpha\beta}
\partial_{\mu}\omega_{\nu}\partial_{\alpha}\rho^{i}_{\beta}
\end{equation}

Using the matrix element (23) and adding the Breit-Wigner formula of the
$\rho$ meson in,
the decay width is obtained
\begin{eqnarray}
\lefteqn{\Gamma=\frac{G^{2}}{128m_{\tau}}
\frac{cos^{2}\theta_{C}}{(2\pi)^{3}}\frac{9}{\pi^{4}g^{4}}\int^{m^{2}_
{\tau}}_{q^{2}_{min}}dq^{2}
{1\over q^{6}}(m^{2}_{\tau}-q^{2})^{2}}\nonumber \\
&&\int^{(\sqrt{q^{2}}-m_{\omega})^{2}}_{4m^{2}_{\pi}}dk^{2}
[(q^{2}+m^{2}_{\omega}-k^{2}
)^{2}-4q^{2}m^{2}_{\omega}]^{{3\over2}}{1\over\pi}
\frac{\sqrt{k^{2}}\Gamma_{\rho}(k^{2})}{(k^{2}-m^{2}_{\rho})^{2}+k^{2}
\Gamma^{2}_{\rho}(k^{2})},
\end{eqnarray}
where \(q^{2}_{min}=2m_{\pi}m_{\tau}+\frac{m_{\tau}m^{2}_{\omega}}{
m_{\tau}-2m_{\pi}}\),
$k^{2}$ is the invariant mass of the two pions and
\begin{eqnarray}
\Gamma_{\rho}(k^{2})=\frac{f^{2}_{\rho\pi\pi}(k^{2})
}{48\pi}{k^{2}\over
m_{\rho}}(1-4{m^{2}_{\pi}\over k^{2}})^{3}, \nonumber \\
f_{\rho\pi\pi}(k^{2})={2\over g}\{1+\frac{k^{2}}{2\pi^{2}f^{2}_{\pi}}
[(1-{2c\over g})^{2}-4\pi^{2}c^{2}]\}.
\end{eqnarray}

The branching ratio is computed to be
\[B=0.16\times10^{-4}.\]
The decay rate is much smaller than the one of $\tau\rightarrow\omega
(\pi\pi)_{non\rho}\nu$. Therefore, $\tau\rightarrow\omega
(\pi\pi)_{non\rho}\nu$
dominates the decay $\tau\rightarrow\omega\pi\pi\nu$.
The distribution is shown in Fig.2.

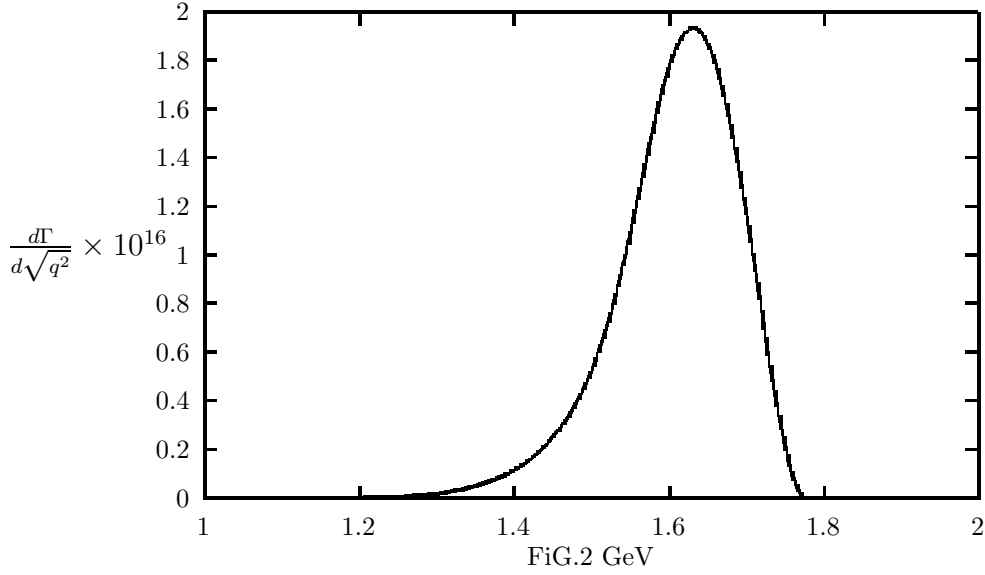
\begin{figure}
\begin{center}

\setlength{\unitlength}{0.240900pt}
\ifx\plotpoint\undefined\newsavebox{\plotpoint}\fi
\sbox{\plotpoint}{\rule[-0.500pt]{1.000pt}{1.000pt}}%
\begin{picture}(1500,900)(0,0)
\font\gnuplot=cmr10 at 10pt
\gnuplot
\sbox{\plotpoint}{\rule[-0.500pt]{1.000pt}{1.000pt}}%
\put(220.0,113.0){\rule[-0.500pt]{292.934pt}{1.000pt}}
\put(220.0,113.0){\rule[-0.500pt]{4.818pt}{1.000pt}}
\put(198,113){\makebox(0,0)[r]{0}}
\put(1416.0,113.0){\rule[-0.500pt]{4.818pt}{1.000pt}}
\put(220.0,189.0){\rule[-0.500pt]{4.818pt}{1.000pt}}
\put(198,189){\makebox(0,0)[r]{0.2}}
\put(1416.0,189.0){\rule[-0.500pt]{4.818pt}{1.000pt}}
\put(220.0,266.0){\rule[-0.500pt]{4.818pt}{1.000pt}}
\put(198,266){\makebox(0,0)[r]{0.4}}
\put(1416.0,266.0){\rule[-0.500pt]{4.818pt}{1.000pt}}
\put(220.0,342.0){\rule[-0.500pt]{4.818pt}{1.000pt}}
\put(198,342){\makebox(0,0)[r]{0.6}}
\put(1416.0,342.0){\rule[-0.500pt]{4.818pt}{1.000pt}}
\put(220.0,419.0){\rule[-0.500pt]{4.818pt}{1.000pt}}
\put(198,419){\makebox(0,0)[r]{0.8}}
\put(1416.0,419.0){\rule[-0.500pt]{4.818pt}{1.000pt}}
\put(220.0,495.0){\rule[-0.500pt]{4.818pt}{1.000pt}}
\put(198,495){\makebox(0,0)[r]{1}}
\put(1416.0,495.0){\rule[-0.500pt]{4.818pt}{1.000pt}}
\put(220.0,571.0){\rule[-0.500pt]{4.818pt}{1.000pt}}
\put(198,571){\makebox(0,0)[r]{1.2}}
\put(1416.0,571.0){\rule[-0.500pt]{4.818pt}{1.000pt}}
\put(220.0,648.0){\rule[-0.500pt]{4.818pt}{1.000pt}}
\put(198,648){\makebox(0,0)[r]{1.4}}
\put(1416.0,648.0){\rule[-0.500pt]{4.818pt}{1.000pt}}
\put(220.0,724.0){\rule[-0.500pt]{4.818pt}{1.000pt}}
\put(198,724){\makebox(0,0)[r]{1.6}}
\put(1416.0,724.0){\rule[-0.500pt]{4.818pt}{1.000pt}}
\put(220.0,801.0){\rule[-0.500pt]{4.818pt}{1.000pt}}
\put(198,801){\makebox(0,0)[r]{1.8}}
\put(1416.0,801.0){\rule[-0.500pt]{4.818pt}{1.000pt}}
\put(220.0,877.0){\rule[-0.500pt]{4.818pt}{1.000pt}}
\put(198,877){\makebox(0,0)[r]{2}}
\put(1416.0,877.0){\rule[-0.500pt]{4.818pt}{1.000pt}}
\put(220.0,113.0){\rule[-0.500pt]{1.000pt}{4.818pt}}
\put(220,68){\makebox(0,0){1}}
\put(220.0,857.0){\rule[-0.500pt]{1.000pt}{4.818pt}}
\put(463.0,113.0){\rule[-0.500pt]{1.000pt}{4.818pt}}
\put(463,68){\makebox(0,0){1.2}}
\put(463.0,857.0){\rule[-0.500pt]{1.000pt}{4.818pt}}
\put(706.0,113.0){\rule[-0.500pt]{1.000pt}{4.818pt}}
\put(706,68){\makebox(0,0){1.4}}
\put(706.0,857.0){\rule[-0.500pt]{1.000pt}{4.818pt}}
\put(950.0,113.0){\rule[-0.500pt]{1.000pt}{4.818pt}}
\put(950,68){\makebox(0,0){1.6}}
\put(950.0,857.0){\rule[-0.500pt]{1.000pt}{4.818pt}}
\put(1193.0,113.0){\rule[-0.500pt]{1.000pt}{4.818pt}}
\put(1193,68){\makebox(0,0){1.8}}
\put(1193.0,857.0){\rule[-0.500pt]{1.000pt}{4.818pt}}
\put(1436.0,113.0){\rule[-0.500pt]{1.000pt}{4.818pt}}
\put(1436,68){\makebox(0,0){2}}
\put(1436.0,857.0){\rule[-0.500pt]{1.000pt}{4.818pt}}
\put(220.0,113.0){\rule[-0.500pt]{292.934pt}{1.000pt}}
\put(1436.0,113.0){\rule[-0.500pt]{1.000pt}{184.048pt}}
\put(220.0,877.0){\rule[-0.500pt]{292.934pt}{1.000pt}}
\put(45,495){\makebox(0,0){${d\Gamma\over d\sqrt{q^{2}}}\times 10^{16}$ }}
\put(828,23){\makebox(0,0){FiG.2     GeV}}
\put(220.0,113.0){\rule[-0.500pt]{1.000pt}{184.048pt}}
\put(344,113){\usebox{\plotpoint}}
\put(467,111.42){\rule{2.409pt}{1.000pt}}
\multiput(467.00,110.92)(5.000,1.000){2}{\rule{1.204pt}{1.000pt}}
\put(344.0,113.0){\rule[-0.500pt]{29.631pt}{1.000pt}}
\put(506,112.42){\rule{2.409pt}{1.000pt}}
\multiput(506.00,111.92)(5.000,1.000){2}{\rule{1.204pt}{1.000pt}}
\put(477.0,114.0){\rule[-0.500pt]{6.986pt}{1.000pt}}
\put(535,113.42){\rule{2.168pt}{1.000pt}}
\multiput(535.00,112.92)(4.500,1.000){2}{\rule{1.084pt}{1.000pt}}
\put(544,114.42){\rule{2.168pt}{1.000pt}}
\multiput(544.00,113.92)(4.500,1.000){2}{\rule{1.084pt}{1.000pt}}
\put(516.0,115.0){\rule[-0.500pt]{4.577pt}{1.000pt}}
\put(563,115.42){\rule{2.168pt}{1.000pt}}
\multiput(563.00,114.92)(4.500,1.000){2}{\rule{1.084pt}{1.000pt}}
\put(572,116.42){\rule{2.168pt}{1.000pt}}
\multiput(572.00,115.92)(4.500,1.000){2}{\rule{1.084pt}{1.000pt}}
\put(581,117.42){\rule{2.168pt}{1.000pt}}
\multiput(581.00,116.92)(4.500,1.000){2}{\rule{1.084pt}{1.000pt}}
\put(590,118.42){\rule{2.168pt}{1.000pt}}
\multiput(590.00,117.92)(4.500,1.000){2}{\rule{1.084pt}{1.000pt}}
\put(599,119.92){\rule{2.168pt}{1.000pt}}
\multiput(599.00,118.92)(4.500,2.000){2}{\rule{1.084pt}{1.000pt}}
\put(608,121.92){\rule{2.168pt}{1.000pt}}
\multiput(608.00,120.92)(4.500,2.000){2}{\rule{1.084pt}{1.000pt}}
\put(617,123.42){\rule{2.168pt}{1.000pt}}
\multiput(617.00,122.92)(4.500,1.000){2}{\rule{1.084pt}{1.000pt}}
\put(626,124.92){\rule{2.168pt}{1.000pt}}
\multiput(626.00,123.92)(4.500,2.000){2}{\rule{1.084pt}{1.000pt}}
\put(635,127.42){\rule{2.168pt}{1.000pt}}
\multiput(635.00,125.92)(4.500,3.000){2}{\rule{1.084pt}{1.000pt}}
\put(644,129.92){\rule{1.927pt}{1.000pt}}
\multiput(644.00,128.92)(4.000,2.000){2}{\rule{0.964pt}{1.000pt}}
\put(652,132.42){\rule{2.168pt}{1.000pt}}
\multiput(652.00,130.92)(4.500,3.000){2}{\rule{1.084pt}{1.000pt}}
\put(661,135.42){\rule{2.168pt}{1.000pt}}
\multiput(661.00,133.92)(4.500,3.000){2}{\rule{1.084pt}{1.000pt}}
\put(670,138.42){\rule{1.927pt}{1.000pt}}
\multiput(670.00,136.92)(4.000,3.000){2}{\rule{0.964pt}{1.000pt}}
\put(678,141.92){\rule{2.168pt}{1.000pt}}
\multiput(678.00,139.92)(4.500,4.000){2}{\rule{1.084pt}{1.000pt}}
\put(687,145.92){\rule{2.168pt}{1.000pt}}
\multiput(687.00,143.92)(4.500,4.000){2}{\rule{1.084pt}{1.000pt}}
\multiput(696.00,151.86)(0.320,0.424){2}{\rule{1.850pt}{0.102pt}}
\multiput(696.00,147.92)(4.160,5.000){2}{\rule{0.925pt}{1.000pt}}
\multiput(704.00,156.86)(0.320,0.424){2}{\rule{1.850pt}{0.102pt}}
\multiput(704.00,152.92)(4.160,5.000){2}{\rule{0.925pt}{1.000pt}}
\multiput(712.00,161.86)(0.490,0.424){2}{\rule{2.050pt}{0.102pt}}
\multiput(712.00,157.92)(4.745,5.000){2}{\rule{1.025pt}{1.000pt}}
\multiput(721.00,166.84)(0.476,0.462){4}{\rule{1.583pt}{0.111pt}}
\multiput(721.00,162.92)(4.714,6.000){2}{\rule{0.792pt}{1.000pt}}
\multiput(729.00,172.84)(0.525,0.475){6}{\rule{1.536pt}{0.114pt}}
\multiput(729.00,168.92)(5.813,7.000){2}{\rule{0.768pt}{1.000pt}}
\multiput(738.00,179.84)(0.444,0.475){6}{\rule{1.393pt}{0.114pt}}
\multiput(738.00,175.92)(5.109,7.000){2}{\rule{0.696pt}{1.000pt}}
\multiput(746.00,186.83)(0.402,0.481){8}{\rule{1.250pt}{0.116pt}}
\multiput(746.00,182.92)(5.406,8.000){2}{\rule{0.625pt}{1.000pt}}
\multiput(754.00,194.83)(0.402,0.481){8}{\rule{1.250pt}{0.116pt}}
\multiput(754.00,190.92)(5.406,8.000){2}{\rule{0.625pt}{1.000pt}}
\multiput(763.83,201.00)(0.481,0.539){8}{\rule{0.116pt}{1.500pt}}
\multiput(759.92,201.00)(8.000,6.887){2}{\rule{1.000pt}{0.750pt}}
\multiput(771.83,211.00)(0.485,0.483){10}{\rule{0.117pt}{1.361pt}}
\multiput(767.92,211.00)(9.000,7.175){2}{\rule{1.000pt}{0.681pt}}
\multiput(780.83,221.00)(0.481,0.608){8}{\rule{0.116pt}{1.625pt}}
\multiput(776.92,221.00)(8.000,7.627){2}{\rule{1.000pt}{0.813pt}}
\multiput(788.83,232.00)(0.481,0.677){8}{\rule{0.116pt}{1.750pt}}
\multiput(784.92,232.00)(8.000,8.368){2}{\rule{1.000pt}{0.875pt}}
\multiput(796.83,244.00)(0.481,0.814){8}{\rule{0.116pt}{2.000pt}}
\multiput(792.92,244.00)(8.000,9.849){2}{\rule{1.000pt}{1.000pt}}
\multiput(804.83,258.00)(0.481,0.883){8}{\rule{0.116pt}{2.125pt}}
\multiput(800.92,258.00)(8.000,10.589){2}{\rule{1.000pt}{1.063pt}}
\multiput(812.83,273.00)(0.481,0.951){8}{\rule{0.116pt}{2.250pt}}
\multiput(808.92,273.00)(8.000,11.330){2}{\rule{1.000pt}{1.125pt}}
\multiput(820.83,289.00)(0.481,1.089){8}{\rule{0.116pt}{2.500pt}}
\multiput(816.92,289.00)(8.000,12.811){2}{\rule{1.000pt}{1.250pt}}
\multiput(828.83,307.00)(0.481,1.158){8}{\rule{0.116pt}{2.625pt}}
\multiput(824.92,307.00)(8.000,13.552){2}{\rule{1.000pt}{1.313pt}}
\multiput(836.84,326.00)(0.475,1.583){6}{\rule{0.114pt}{3.393pt}}
\multiput(832.92,326.00)(7.000,14.958){2}{\rule{1.000pt}{1.696pt}}
\multiput(843.83,348.00)(0.481,1.433){8}{\rule{0.116pt}{3.125pt}}
\multiput(839.92,348.00)(8.000,16.514){2}{\rule{1.000pt}{1.563pt}}
\multiput(851.83,371.00)(0.481,1.639){8}{\rule{0.116pt}{3.500pt}}
\multiput(847.92,371.00)(8.000,18.736){2}{\rule{1.000pt}{1.750pt}}
\multiput(859.83,397.00)(0.481,1.776){8}{\rule{0.116pt}{3.750pt}}
\multiput(855.92,397.00)(8.000,20.217){2}{\rule{1.000pt}{1.875pt}}
\multiput(867.84,425.00)(0.475,2.234){6}{\rule{0.114pt}{4.536pt}}
\multiput(863.92,425.00)(7.000,20.586){2}{\rule{1.000pt}{2.268pt}}
\multiput(874.83,455.00)(0.481,2.051){8}{\rule{0.116pt}{4.250pt}}
\multiput(870.92,455.00)(8.000,23.179){2}{\rule{1.000pt}{2.125pt}}
\multiput(882.83,487.00)(0.481,2.189){8}{\rule{0.116pt}{4.500pt}}
\multiput(878.92,487.00)(8.000,24.660){2}{\rule{1.000pt}{2.250pt}}
\multiput(890.84,521.00)(0.475,2.641){6}{\rule{0.114pt}{5.250pt}}
\multiput(886.92,521.00)(7.000,24.103){2}{\rule{1.000pt}{2.625pt}}
\multiput(897.83,556.00)(0.481,2.257){8}{\rule{0.116pt}{4.625pt}}
\multiput(893.92,556.00)(8.000,25.401){2}{\rule{1.000pt}{2.312pt}}
\multiput(905.83,591.00)(0.481,2.326){8}{\rule{0.116pt}{4.750pt}}
\multiput(901.92,591.00)(8.000,26.141){2}{\rule{1.000pt}{2.375pt}}
\multiput(913.84,627.00)(0.475,2.641){6}{\rule{0.114pt}{5.250pt}}
\multiput(909.92,627.00)(7.000,24.103){2}{\rule{1.000pt}{2.625pt}}
\multiput(920.83,662.00)(0.481,2.120){8}{\rule{0.116pt}{4.375pt}}
\multiput(916.92,662.00)(8.000,23.919){2}{\rule{1.000pt}{2.188pt}}
\multiput(928.84,695.00)(0.475,2.397){6}{\rule{0.114pt}{4.821pt}}
\multiput(924.92,695.00)(7.000,21.993){2}{\rule{1.000pt}{2.411pt}}
\multiput(935.84,727.00)(0.475,2.153){6}{\rule{0.114pt}{4.393pt}}
\multiput(931.92,727.00)(7.000,19.882){2}{\rule{1.000pt}{2.196pt}}
\multiput(942.83,756.00)(0.481,1.639){8}{\rule{0.116pt}{3.500pt}}
\multiput(938.92,756.00)(8.000,18.736){2}{\rule{1.000pt}{1.750pt}}
\multiput(950.84,782.00)(0.475,1.583){6}{\rule{0.114pt}{3.393pt}}
\multiput(946.92,782.00)(7.000,14.958){2}{\rule{1.000pt}{1.696pt}}
\multiput(957.83,804.00)(0.481,1.158){8}{\rule{0.116pt}{2.625pt}}
\multiput(953.92,804.00)(8.000,13.552){2}{\rule{1.000pt}{1.313pt}}
\multiput(965.84,823.00)(0.475,0.932){6}{\rule{0.114pt}{2.250pt}}
\multiput(961.92,823.00)(7.000,9.330){2}{\rule{1.000pt}{1.125pt}}
\multiput(972.84,837.00)(0.475,0.525){6}{\rule{0.114pt}{1.536pt}}
\multiput(968.92,837.00)(7.000,5.813){2}{\rule{1.000pt}{0.768pt}}
\multiput(978.00,847.86)(0.320,0.424){2}{\rule{1.850pt}{0.102pt}}
\multiput(978.00,843.92)(4.160,5.000){2}{\rule{0.925pt}{1.000pt}}
\put(553.0,117.0){\rule[-0.500pt]{2.409pt}{1.000pt}}
\put(993,846.92){\rule{1.686pt}{1.000pt}}
\multiput(993.00,848.92)(3.500,-4.000){2}{\rule{0.843pt}{1.000pt}}
\multiput(1001.84,840.03)(0.475,-0.606){6}{\rule{0.114pt}{1.679pt}}
\multiput(997.92,843.52)(7.000,-6.516){2}{\rule{1.000pt}{0.839pt}}
\multiput(1008.84,827.66)(0.475,-0.932){6}{\rule{0.114pt}{2.250pt}}
\multiput(1004.92,832.33)(7.000,-9.330){2}{\rule{1.000pt}{1.125pt}}
\multiput(1015.83,812.10)(0.481,-1.158){8}{\rule{0.116pt}{2.625pt}}
\multiput(1011.92,817.55)(8.000,-13.552){2}{\rule{1.000pt}{1.313pt}}
\multiput(1023.84,788.73)(0.475,-1.746){6}{\rule{0.114pt}{3.679pt}}
\multiput(1019.92,796.36)(7.000,-16.365){2}{\rule{1.000pt}{1.839pt}}
\multiput(1030.84,762.95)(0.475,-1.990){6}{\rule{0.114pt}{4.107pt}}
\multiput(1026.92,771.48)(7.000,-18.475){2}{\rule{1.000pt}{2.054pt}}
\multiput(1037.84,734.17)(0.475,-2.234){6}{\rule{0.114pt}{4.536pt}}
\multiput(1033.92,743.59)(7.000,-20.586){2}{\rule{1.000pt}{2.268pt}}
\multiput(1044.84,701.80)(0.475,-2.560){6}{\rule{0.114pt}{5.107pt}}
\multiput(1040.92,712.40)(7.000,-23.400){2}{\rule{1.000pt}{2.554pt}}
\multiput(1051.84,666.61)(0.475,-2.723){6}{\rule{0.114pt}{5.393pt}}
\multiput(1047.92,677.81)(7.000,-24.807){2}{\rule{1.000pt}{2.696pt}}
\multiput(1058.84,628.24)(0.475,-3.048){6}{\rule{0.114pt}{5.964pt}}
\multiput(1054.92,640.62)(7.000,-27.621){2}{\rule{1.000pt}{2.982pt}}
\multiput(1065.84,587.65)(0.475,-3.130){6}{\rule{0.114pt}{6.107pt}}
\multiput(1061.92,600.32)(7.000,-28.324){2}{\rule{1.000pt}{3.054pt}}
\multiput(1072.84,544.87)(0.475,-3.374){6}{\rule{0.114pt}{6.536pt}}
\multiput(1068.92,558.43)(7.000,-30.435){2}{\rule{1.000pt}{3.268pt}}
\multiput(1079.84,500.87)(0.475,-3.374){6}{\rule{0.114pt}{6.536pt}}
\multiput(1075.92,514.43)(7.000,-30.435){2}{\rule{1.000pt}{3.268pt}}
\multiput(1086.84,455.68)(0.475,-3.537){6}{\rule{0.114pt}{6.821pt}}
\multiput(1082.92,469.84)(7.000,-31.842){2}{\rule{1.000pt}{3.411pt}}
\multiput(1093.84,410.87)(0.475,-3.374){6}{\rule{0.114pt}{6.536pt}}
\multiput(1089.92,424.43)(7.000,-30.435){2}{\rule{1.000pt}{3.268pt}}
\multiput(1100.84,362.52)(0.462,-4.174){4}{\rule{0.111pt}{7.583pt}}
\multiput(1096.92,378.26)(6.000,-28.260){2}{\rule{1.000pt}{3.792pt}}
\multiput(1106.84,324.06)(0.475,-3.211){6}{\rule{0.114pt}{6.250pt}}
\multiput(1102.92,337.03)(7.000,-29.028){2}{\rule{1.000pt}{3.125pt}}
\multiput(1113.84,283.83)(0.475,-2.967){6}{\rule{0.114pt}{5.821pt}}
\multiput(1109.92,295.92)(7.000,-26.917){2}{\rule{1.000pt}{2.911pt}}
\multiput(1120.84,246.02)(0.475,-2.804){6}{\rule{0.114pt}{5.536pt}}
\multiput(1116.92,257.51)(7.000,-25.510){2}{\rule{1.000pt}{2.768pt}}
\multiput(1127.84,211.39)(0.475,-2.479){6}{\rule{0.114pt}{4.964pt}}
\multiput(1123.92,221.70)(7.000,-22.696){2}{\rule{1.000pt}{2.482pt}}
\multiput(1134.84,177.90)(0.462,-2.633){4}{\rule{0.111pt}{5.083pt}}
\multiput(1130.92,188.45)(6.000,-18.449){2}{\rule{1.000pt}{2.542pt}}
\multiput(1140.84,154.73)(0.475,-1.746){6}{\rule{0.114pt}{3.679pt}}
\multiput(1136.92,162.36)(7.000,-16.365){2}{\rule{1.000pt}{1.839pt}}
\multiput(1147.84,134.29)(0.475,-1.258){6}{\rule{0.114pt}{2.821pt}}
\multiput(1143.92,140.14)(7.000,-12.144){2}{\rule{1.000pt}{1.411pt}}
\multiput(1154.84,120.44)(0.475,-0.688){6}{\rule{0.114pt}{1.821pt}}
\multiput(1150.92,124.22)(7.000,-7.220){2}{\rule{1.000pt}{0.911pt}}
\put(986.0,851.0){\rule[-0.500pt]{1.686pt}{1.000pt}}
\end{picture}

\end{center}
\end{figure}

The nonconservation of the quark axial-vector currents
found in the matrix
element(13,23) show that there is anomaly in the PCAC.
The Adler-Bell-Jackiw anomaly
\begin{equation}
\partial^{\mu}\bar{\psi}\tau_{i}\gamma_{\mu}\gamma_{5}\psi=
\frac{\alpha}{4\pi}\varepsilon^{\mu\nu\alpha\beta}F_{\mu\nu}
F_{\alpha\beta}\delta_{3i}
\end{equation}
is well known.
However, this anomaly is the one caused by electromagnetic interaction.
The abnormal terms(14,24) are from strong interaction.
We claim the existence of the strong anomaly in PCAC.
Taking the abnormal term(23) as an example of the strong anomaly of
the PCAC, the PCAC with strong anomaly is written as
\begin{equation}
\partial^{\mu}\bar{\psi}\tau_{i}\gamma_{\mu}\gamma_{5}\psi=
-m^{2}_{\pi}f_{\pi}\pi_{i}+\frac{N_{C}}{\pi^{2}g^{2}}
\varepsilon^{\mu\nu\alpha\beta}\partial_{\mu}\omega_{\nu}
\partial_{\alpha}\rho^{i}_{\beta}.
\end{equation}
The abnormal term of Eq.(28) is caused by the vertex(20) which
originates in the WZW anomaly.
Why this vertex causes the anomaly of the PCAC? The reason is
presented as below.
As pointed in Ref.[6], the conservation of the quark axial-vector
currents(in the chiral limit) is obtained from the cancellations
between pion exchange, $a_{1}$ exchange, and others.
In the Lagrangian of the
WZW anomaly[4,5,7] there is no coupling like
\[\sim\varepsilon^{\mu\nu\alpha\beta}\partial_{\mu}\omega_{\nu}
\rho^{i}_{\alpha}a^{i}_{\beta}.\]
Therefore, the pion exchange provided by the vertex(20) cannot be
canceled out. The lack of the cancellation leads to the
nonconservation of the quark axial-vector currents in the chiral limit.

We use the effective chiral theory[5] to illustrate the existence
of the strong anomaly of PCAC(28).
The problem is similar to the one treated by
Adler-Bell-Jackiw[1].
First it is necessary to show that in the chiral limit, if just
applying the equations of motion to the quark
axial-vector current, it is found that the currents are conserved in
the chiral limit.
The equations of motion
are derived from the Lagrangian(6)(to avoid the $U(1)$ problem taking
off the $\eta$ field in this part of the discussion)
\begin{eqnarray}
\lefteqn{\partial^{\mu}\bar{\psi}\gamma_{\mu}=-i\bar{\psi}(
\gamma_{\mu}v^{\mu}+\gamma_{\mu}\gamma_{5}a^{\mu}-mu-M)
,}\nonumber \\
&&\gamma_{\mu}\partial^{\mu}\psi=i(\gamma_{\mu}v^{\mu}+\gamma_{\mu}
\gamma_{5}a^{\mu}-mu-M)\psi,\nonumber \\
&&\rho^{i}_{\mu}=-{1\over m^{2}_{0}}\bar{\psi}\tau_{i}\gamma_{\mu}
\psi\;\;\;
a^{i}_{\mu}=-{1\over m^{2}_{0}}\bar{\psi}\tau_{i}\gamma_{\mu}
\gamma_{5}\psi,\nonumber \\
&&\Pi_{i}={i\sigma}\frac{
\bar{\psi}\tau_{i}\gamma_{5}\psi}{\bar{\psi}\psi}
,
\end{eqnarray}
where \(u=\sigma+i\gamma_{5}\tau\cdot \Pi\), \(\sigma^{2}=1-\Pi^{2}
\).
Using all these equations(29), to the leading order of the quark
mass it is proved
\begin{equation}
\partial^{\mu}\bar{\psi}\tau_{i}\gamma_{\mu}\gamma_{5}\psi=
-m^{2}_{\pi}f_{\pi}\pi^{i}.
\end{equation}
According to the Adler-Bell-Jackiw[1]
the anomaly does not come from the classic equation of motion,
instead, from renormalization of the quark triangle diagrams.
Adding the photon field to the Lagrangian(6), the vector part
of the Lagrangian is written as
\begin{equation}
{\cal L}=\bar{\psi}(x)\{i\gamma\cdot\partial+\gamma^{\mu}(
{1\over g}\tau_{3}\rho^{0}_{\mu}+{1\over2}e\tau_{3}A_{\mu}
+{1\over g}\omega_{\mu}+{1\over6}eA_{\mu})
\}\psi(x)
\end{equation}
This part of the Lagrangian shows that the $\rho$, $\omega$ and the
photon fields are in symmetric positions
which lead to the VMD[5].
As pointed out by Sakurai[8],
the substitutions
\begin{equation}
\rho^{i}\rightarrow{1\over2}egA,\;\;\;
\omega\rightarrow{1\over6}egA
\end{equation}
revealed from Eq.(31) are essential to obtain VMD.
Taking $\rho$ and $\omega$ fields as external fields and
calculating the quark triangle diagrams as done by Adler-Bell-Jackiw,
the strong anomaly of PCAC is derived as the one shown in Eq.(28).
It is equivalent to say that
this anomaly(28) can be found by using the
substitution
\[Tr\tau^{3}Q^{2}\partial_{\mu}A_{\nu}\partial_{\alpha}A_{\beta}
\rightarrow 2Tr\tau_{3}\tau_{3}\partial_{\mu}\omega_{\nu}
\partial_{\alpha}\rho^{3}_{\beta} \]
in Eq.(27).

The abnormal term of Eq.(28) can be written as the divergence
of the current
\begin{equation}
\frac{N_{C}}{2\pi^{2}g^{2}}\varepsilon^{\mu\nu\alpha\beta}
\{\omega_{\nu}\partial_{\alpha}\rho^{i}_{\beta}+
\rho^{i}_{\nu}\partial_{\alpha}\omega_{\beta}\}.
\end{equation}
Therefore, a question is raised that whether this current is part of
the quark axial-vector current. If so, there is no strong interaction
anomaly in PCAC. A direct proof is necessary.
The method using the quark operator to bosonize the quark currents
is presented in Ref.[5]. The couplings between $\omega$ meson and
others obtained by using this method is the same as the one
derived from the WZW Lagrangian[4,7].
Following this method,
we have
\begin{eqnarray}
\lefteqn{<\bar{\psi}\tau_{i}\gamma_{\mu}\gamma_{5}\psi>
=-\frac{iN_{C}}{(2\pi)^{D}}\int d^{D}p Tr\tau_{i}\gamma_{\mu}
\gamma_{5}s_{F}(x,p),}\\
&&s_{F}(x,p)=s^{0}_{F}(p)\sum^{\infty}_{n=0}(-i)^{n}\{\gamma^{\mu}
D_{\mu}s^{0}_{F}(p)\}^{n},\\
&&D_{\mu}=\partial_{\mu}-iv_{\mu}-ia_{\mu}\gamma_{5},\\
&&s^{0}_{F}(p)=-\frac{\gamma^{\mu}p_{\mu}-m\hat{u}}{p^{2}-m^{2}},\\
&&\hat{u}=exp^{-i\gamma_{5}(\tau^{i}\pi^{i}+\eta)}.
\end{eqnarray}
The effective chiral theory of mesons[5] is a theory of mesons
at low energies and the derivative expansion has been exploited.
We are only interested in the terms associated with the
antisymmetric tensor $\varepsilon^{\mu\nu\alpha\beta}$. The leading
terms are from \(n=3\)
\begin{equation}
<\bar{\psi}\tau_{i}\gamma_{\mu}\gamma_{5}\psi>
=\frac{N_{C}}{(2\pi)^{D}}\int d^{D}p\frac{1}{(p^{2}-m^{2})^{4}}
Tr\tau_{i}\gamma_{\mu}\gamma_{5}(\gamma\cdot p-m)\gamma\cdot
D(\gamma\cdot p-m)\gamma\cdot D(\gamma\cdot p-m)
\gamma\cdot D(\gamma\cdot p-m).
\end{equation}
We are looking for the terms containing one $\omega$ and one $\rho$
field only.
The derivation shows that all nonzero terms are cancelled out.
Therefore, the term(33) is not included in the quark axial-vector
current. This conclusion is consistent with the fact that in the
WZW Lagrangian there is no coupling between $\omega$, $\rho$, and
$a_{1}$ fields. The explanation is following.
In the effective chiral theory of mesons[5]
the couplings between $a_{1}$ fields and others are obtained
from part of the Lagrangian(6)
\[a^{i}_{\mu}<\bar{\psi}\tau_{i}\gamma_{\mu}\gamma_{5}\psi>.\]
As obtained above, in the effective axial-vector currents(39) there is
no terms containing both $\omega$ and $\rho$ fields only. Therefore,
there is no $a_{1}\omega\rho$ coupling.

We emphasize on that the  PCAC with strong anomaly(28) is model
independent.
The vertex(20) is derived from the WZW Lagrangian[4,5,7]
and is very general. The term $-{g_{W}\over4}cos\theta_{C}f_{\pi}
A^{i}_{\mu}\partial^{\mu}\pi^{i}$ used to derived the matrix element
(23) is independent of any model.

The vertex(20) has been well tested.
In Ref.[5] it has been used to derive
the amplitude of $\pi^{0}\rightarrow2\gamma$ obtained by Adler-Bell
-Jackiw triangle anomaly. The decay rates of $\omega
\rightarrow\pi\gamma$ and $\rho\rightarrow\pi\gamma$ are via VMD
calculated
by using this vertex and theoretical results are in good agreements
with data. It is also shown in Ref.[5] that this vertex is responsible
for the decay $\omega3\pi$.

On the other hand, using the substitutions(32) in Eq.(28) we derive

\begin{eqnarray}
\lefteqn{\partial^{\mu}\bar{\psi}\tau_{i}\gamma_{\mu}\gamma_{5}\psi=
-m^{2}_{\pi}f_{\pi}\pi_{i}+\frac{N_{C}}{\pi^{2}g^{2}}
\varepsilon^{\mu\nu\alpha\beta}\partial_{\mu}\omega_{\nu}
\partial_{\alpha}\rho^{i}_{\beta}
+\frac{\alpha}{4\pi}\varepsilon^{\mu\nu\alpha\beta}F_{\mu\nu}
F_{\alpha\beta}\delta_{3i}}\nonumber \\
&&+\frac{e}{4\pi^{2}g}\varepsilon^{\mu\nu\alpha\beta}F_{\mu\nu}
\partial_{\alpha}\rho^{i}_{\beta}+\frac{3e}{4\pi^{2}g}
\varepsilon^{\mu\nu\alpha\beta}\partial_{\mu}\omega_{\nu}F_{\alpha
\beta}\delta_{3i}.
\end{eqnarray}
It is well known that the amplitude of $\pi^{0}\rightarrow2\gamma$
is derived from the third term of the Eq.(41) under a soft pion
approximation. In the same way, the amplitudes of $\omega\rightarrow
\pi\gamma$ and $\rho\rightarrow\pi\gamma$ are obtained from the
fourth and the fifth term of Eq.(40)
\begin{equation}
{\cal M}(\omega\rightarrow\pi\gamma)=\frac{3e}{2\pi^{2}g}\varepsilon
^{\mu\nu\alpha\beta}p_{\mu}k_{\alpha}\epsilon_{\nu}(p)\epsilon^{*}
_{\beta}(k),
\end{equation}
where k and p are momentum of $\omega$ and photon respectively.
\begin{equation}
{\cal M}(\rho\rightarrow\pi\gamma)=\frac{e}{2\pi^{2}g}\varepsilon
^{\mu\nu\alpha\beta}p_{\mu}k_{\alpha}\epsilon_{\nu}(p)\epsilon^{*}
_{\beta}(k),
\end{equation}
The decay widths are
\begin{eqnarray}
\Gamma(\omega\rightarrow\pi\gamma)=\frac{3\alpha}{32\pi^{4}g^{2}}
\frac{m^{3}_{\omega}}{f^{2}_{\pi}}(1-{m^{2}_{\pi}\over m^{2}_{\omega}
})^{3}=583keV,\\
\Gamma(\rho\rightarrow\pi\gamma)=\frac{\alpha}{96\pi^{4}g^{2}}
\frac{m^{3}_{\rho}}{f^{2}_{\pi}}(1-{m^{2}_{\pi}\over m^{2}_{\rho}
})^{3}=61keV.
\end{eqnarray}
The data are $717(1\pm0.07)$keV and $67.8(1\pm0.12)$keV
[10] respectively.

It is known that $\omega\rightarrow\rho\pi$ is dominant in the decay
$\omega\rightarrow3\pi$. In the manner of the calculations done
above, the PCAC with anomaly(28) is used to
calculate the decay rate of $\omega\rightarrow3\pi$
\begin{eqnarray}
\lefteqn{\Gamma(\omega\rightarrow3\pi)=
{1\over384m^{3}_{\omega}}{1\over(2\pi)
^{3}}\int dq^{2}_{1}dq^{2}_{2}(m^{2}_{\omega}-q^{2}_{1}-q^{2}_{2})
\{
(m^{2}_{\omega}-q^{2}_{1})(m^{2}_{\omega}-q^{2}_{2}) }\nonumber \\
&&-m^{2}_{\omega}
(q^{2}_{1}+q^{2}_{2}-m^{2}_{\omega})\}
\{\frac{f^{2}_{\rho\pi\pi}(q^{2}_{1})}{q^{2}_{1}-m^{2}_{\rho}}
+\frac{f^{2}_{\rho\pi\pi}(q^{2}_{2})}{q^{2}_{2}-m^{2}_{\rho}}
+\frac{f^{2}_{\rho\pi\pi}(q^{2}_{3})}{q^{2}_{3}-m^{2}_{\rho}}
\}^{2}.
\end{eqnarray}
In Eq.(45) the amplitude of $\omega\rightarrow3\pi$ is determined
in the chiral limit. The numerical result is
\[\Gamma(\omega\rightarrow3\pi)=7.7MeV.\]
The data is 7.49(1$\pm$0.02)MeV[10].

There are more terms on the right hand side of the Eq.(40). For
example, the term(14) should be added to the right hand side of Eq.(28).
In the same way obtaining the term(24),
the vertex
\begin{equation}
{\cal L}^{f_{1}a_{1}\pi}=\frac{1}{\pi^{2}f_{\pi}}f^{2}_{a}
\varepsilon^{\mu\nu\alpha\beta}f_{\mu}\partial_{\nu}
\pi^{i}\partial_{\alpha}a^{i}_{\beta}
\end{equation}
used to calculate the decay width of $\tau\rightarrow f\pi\nu$ in
Ref.[6] results a term
\begin{equation}
-{1\over\pi^{2}}f^{2}_{a}\varepsilon^{\mu\nu\alpha\beta}\partial
_{\mu}f_{\nu}\partial_{\alpha}a^{i}_{\beta}
\end{equation}
which should be part of the strong anomaly of the PCAC.
Now the PCAC with strong anomaly
takes the form
\begin{eqnarray}
\lefteqn{\partial^{\mu}\bar{\psi}\tau_{i}\gamma_{\mu}\gamma_{5}\psi=
-m^{2}_{\pi}f_{\pi}\pi_{i}
+\frac{N_{C}}{\pi^{2}g^{2}}\varepsilon^{\mu\nu\alpha\beta}
\partial_{\mu}\omega_{\nu}\partial_{\alpha}\rho^{i}_{\beta}}\nonumber \\
&&+\frac{6}{\pi^{2}gf^{2}_{\pi}}{2c\over g}(1-{2c\over g})\varepsilon
^{\mu\nu\alpha\beta}\epsilon_{ijk}\partial_{\nu}\omega_{\mu}\partial
_{\alpha}\pi^{j}\partial_{\beta}\pi^{k}
-{1\over\pi^{2}}f^{2}_{a}\varepsilon^{\mu\nu\alpha\beta}\partial
_{\mu}f_{\nu}\partial_{\alpha}a^{i}_{\beta}.
\end{eqnarray}

Obviously, there are much more terms for the strong anomaly of the PCAC.
The method deriving those abnormal terms is the same as the one used
to obtained (14) and (24). By taking away the factor $-{g_{W}\over 4}
cos\theta_{C}A^{i}_{\mu}$ from ${\cal L}^{A}$(3),
the axial-vector currents are
obtained. Combining these currents with proper vertices of mesons,
the nonconservetive currents, if they exist, could be found.

It is
necessary to emphasize that the decay $\tau\rightarrow\omega\rho\nu$
provides a direct evidence of the strong anomaly of the PCAC.
Therefore, the measurements of the decay rate
and the distribution of the decay rate versus the invariant mass of
$\omega\rho$ will evidence the existence of the strong anomaly in
the PCAC.

The discussion of the strong anomaly of the PCAC in two flavor case can be
extended to three flavors.
The decay $\tau\rightarrow K^{*}\rho\nu$ and
$K^{*}\omega\nu$ are related to the anomaly of the PCAC of \(\Delta s=1\).
Both the vector and axial-vector currents contribute to these decays.
The vertices contributing to the matrix elements of the vector currents
come from the term
\begin{equation}
-{1\over8}Trv_{\mu\nu}v^{\mu\nu}
\end{equation}
of the effective Lagrangian of mesons obtained from the Lagrangian(6)
in Ref.[5], where
\[v_{\mu\nu}=\partial_{\mu}v_{\nu}-\partial_{\nu}v_{\mu}-{i\over g}
[v_{\mu}, v_{\nu}]-{i\over g}[a_{\mu}, a_{\nu}],\]
\[v_{\mu}=\tau^{i}\rho^{i}_{\mu}+({2\over3}+{1\over\sqrt{3}}\lambda_{8}
)\omega_{\mu}+\lambda_{a}K^{a}_{\mu}
+({1\over3}-{1\over\sqrt{3}}\lambda_{8})\phi_{\mu},\]
\[a_{\mu}=\tau^{i}a^{i}_{\mu}+({2\over3}+{1\over\sqrt{3}}\lambda_{8}
)f_{\mu}+\lambda_{a}K^{a}_{\mu}
+({1\over3}-{1\over\sqrt{3}}\lambda_{8})f_{1s\mu},\]
where \(a=4,5,6,7\).
There are additional normalization factors for $\phi$, $a_{1}$, f, and
$f_{1s}$, which can be found in Ref.[5]. The vertices ${\cal L}^{K^{*}
\bar{K}^{*}v}$ is derived from Eq.(49)
\begin{eqnarray}
\lefteqn{{\cal L}^{K^{*}\bar{K}^{*}v}=\frac{\sqrt{3}i}{g}\{(\partial_
{\mu}v^{8}_{\nu}-\partial_{\nu}v^{8}_{\mu})(K^{-\mu}K^{+\nu}+\bar{K}
^{0\mu}K^{0\nu})+}\nonumber \\
&&v^{8\nu}[
(\partial_{\mu}K^{-}_{\nu}-\partial_{\nu}K^{-}_{\mu})
K^{+\mu}-
(\partial_{\mu}K^{+}_{\nu}-\partial_{\nu}K^{+}_{\mu})
K^{-\mu}+
(\partial_{\mu}\bar{K}^{0}_{\nu}-\partial_{\nu}\bar{K}^{0}_{\mu})
K^{0\mu}-
(\partial_{\mu}K^{0}_{\nu}-\partial_{\nu}K^{0}_{\mu})
\bar{K}^{0\mu}]\}\nonumber \\
&&+{i\over g}\{\sqrt{2}
(\partial_{\mu}\rho^{+}_{\nu}-\partial_{\nu}\rho^{+}
_{\mu})K^{-\mu}K^{0\nu}-\sqrt{2}
(\partial_{\mu}\rho^{-}_{\nu}-\partial_{\nu}\rho^{-}
_{\mu})K^{+\mu}\bar{K}^{0\nu}+
(\partial_{\mu}\rho^{0}_{\nu}-\partial_{\nu}\rho^{0}
_{\mu})(K^{-\mu}K^{+\nu}+\bar{K}^{0\mu}K^{0\nu})\nonumber \\
&&+\sqrt{2}\rho^{+\nu}[(\partial_{\mu}K^{-}_{\nu}-\partial_{\nu}K^{-}
_{\mu})K^{0\mu}-
(\partial_{\mu}K^{0}_{\nu}-\partial_{\nu}K^{0}_{\mu})K
^{-\mu}]\nonumber \\
&&-\sqrt{2}\rho^{-\nu}[(\partial_{\mu}K^{+}_{\nu}-\partial
_{\nu}K^{+}_{\mu})\bar{K}^{0\mu}-
(\partial_{\mu}\bar{K}^{0}_{\nu}-\partial_{\nu}\bar{K}^{0}_{\mu})K
^{+\mu}]\nonumber \\
&&+\rho^{0\nu}[(\partial_{\mu}K^{0}_{\nu}-\partial
_{\nu}K^{0}_{\mu})\bar{K}^{+\mu}-
(\partial_{\mu}K^{+}_{\nu}-\partial_{\nu}K^{+}_{\mu})K
^{-\mu}\nonumber \\
&&-(\partial_{\mu}\bar{K}^{0}_{\nu}-\partial_{\nu}\bar{K}^{0}_{\mu})K
^{0\mu}
+(\partial_{\mu}K^{0}_{\nu}-\partial
_{\nu}K^{0}_{\mu})\bar{K}^{-\mu}]\},
\end{eqnarray}
where
\[v^{8}={1\over\sqrt{3}}\omega-{2\over\sqrt{3}}\phi.\]
Using the substitutions[5]
\begin{equation}
\rho^{0}\rightarrow {1\over2}egA,\;\;\;\omega\rightarrow{1\over6}egA,
\;\;\;\phi\rightarrow-{1\over3\sqrt{2}}egA,
\end{equation}
it is proved that the charges of $K^{*+-0}$ are $+1$, $-1$, and 0
respectively, where A is the photon field. It is interesting to notice
that the vertices(50) are from the nonabelian nature of the vector
meson fields.
The vertices ${\cal L}^{K^{*}\bar{K}^{*}\rho}$ and ${\cal L}^{K^{*}
\bar{K}^{*}\omega}$ are found from Eq.(50).
Using ${\cal L}^{V}$(1), it is obtained
\begin{eqnarray}
\lefteqn{<\rho^{0}K^{*-}|\bar{\psi}\lambda_{+}\gamma_{\mu}\psi|0>=
{1\over2\sqrt{4E_{1}E_{2}}}sin\theta_{C}\frac{-m^{2}_{K^{*}}
+i\sqrt{q^{2}}\Gamma_{K^{*}}(q^{2})}{q^{2}-m^{2}_{K^{*}}+
i\sqrt{q^{2}}
\Gamma_{K^{*}}q^{2})}}\nonumber \\
&&(\frac{q_{\mu}q_{\nu}}{q^{2}}-g_{\mu\nu})\{(k-p)^{\nu}\epsilon_{
\sigma}(k)\epsilon^{\sigma}(p)+2p^{\sigma}\epsilon^{\nu}(p)\epsilon
_{\sigma}(k)
-2k^{\sigma}\epsilon^{\nu}(k)\epsilon_{\sigma}(p)\},
\end{eqnarray}
where k and p are the momentum of $\rho$ and $K^{*}$ respectively,
\(q=k+p\),
\begin{equation}
\Gamma_{K^{*}}(q^{2})=
\frac{f^{2}_{\rho\pi\pi}(q^{2})}{8\pi\sqrt{q^{2}}
m_{K^{*}}}\{{1\over4q^{2}}(q^{2}+m^{2}_{K}+m^{2}_{\pi})^{2}-m^{2}_{K}\}
^{{3\over2}}.
\end{equation}

The vertices contributing to the matrix element of the quark
axial-vector currents are from the WZW anomaly[5]
\begin{equation}
{\cal L}^{K^{*}Kv}=
-\frac{N_{C}}{\pi^{2}g^{2}f_{\pi}}\varepsilon^{
\mu\nu\alpha\beta}d_{abc}K^{a}_{\mu}\partial_{\nu}v^{c}_{\alpha}
\partial_{\beta}K^{b}
-\frac{N_{C}}{\pi^{2}g^{2}f_{\pi}}\varepsilon^{
\mu\nu\alpha\beta}K^{a}_{\mu}\partial_{\nu}v_{\alpha}
\partial_{\beta}K^{a},
\end{equation}
where $v_{\alpha}$ is the singlet.
The vertices $K^{*}K\rho$ and $K^{*}K\omega$ are found
from Eq.(54).
Using these vertices and ${\cal L}^{A}$(3), the matrix element of the
axial-vector currents is obtained
\begin{eqnarray}
\lefteqn{<K^{*-}\rho^{0}|\bar{\psi}\lambda_{+}
\gamma_{\mu}\gamma_{5}\psi|0>=
sin\theta_{C}<K^{*-}\rho^{0}|f_{K}\partial_{\mu}K^{+}|0>=}\nonumber \\
&&\frac{i}{\sqrt{4E_{1}E_{2}}}sin\theta_{C}\frac{N_{C}}{2\pi^{2}g^{2}}
\frac{q_{\mu}}{q^{2}}
\varepsilon^{\lambda\nu\alpha\beta}k_{\nu}\epsilon_{\alpha}(k)
\epsilon_{\lambda}(p)q_{\beta}.
\end{eqnarray}
In the chiral limit, \(f_{K}=f_{\pi}\).
Obviously, the axial-vector current is not conserved in the limit of
\(m_{q}=0\). The divergence of the term contributing to this matrix
element is written as
\begin{equation}
-\frac{N_{C}}{2\pi^{2}g^{2}}\varepsilon^{\mu\nu\alpha
\beta}\partial_{\mu}K^{*-}_{\nu}\partial_{\alpha}\rho^{0}_{\beta}.
\end{equation}

Adding the Breit-Wigner distribution of the $\rho$ meson and
using the matrix elements(52,55), the distributions of the decay $\tau
\rightarrow K^{*-}\rho^{0}\nu$ versus
the invariant mass of $K^{*}\rho$
are obtained
\begin{eqnarray}
\lefteqn{\frac{d\Gamma^{V}}{dq^{2}}=\frac{G^{2}}{(2\pi)^{3}}
\frac{sin^{2}_
{\theta_{C}}}{32m^{3}_{\tau}q^{4}}(m^{2}_{\tau}-q^{2})^{2}
\int^{(\sqrt{q^{2}}-m_{K^{*}})^{2}}_{4m^{2}_{\pi}}dk^{2}
\{{1\over4}(q^{2}+k^{2}-m^{2}_{K^{*}})^{2}
-q^{2}k^{2}\}^{{3\over2}}
\frac{m^{2}_{\tau}+2q^{2}}{q^{2}k^{2}m^{2}_{K^{*}}}}\nonumber \\
&&{1\over\pi}\frac{\sqrt{k^{2}}\Gamma_{\rho}(k^{2})}
{(k^{2}-m^{2}_{\rho})^{2}
+k^{2}\Gamma^{2}_{\rho}(k^{2})}
\frac{m^{4}_{K^{*}}+q^{2}\Gamma^{2}_{K^{*}}(q^{2})}{(q^{2}-m^{2}
_{K^{*}})^{2}+q^{2}\Gamma^{2}_{K^{*}}(q^{2})}\nonumber \\
&&\{q^{2}(k^{2}+m^{2}_{K^{*}})+k^{2}m^{2}_{K^{*}}
+{1\over12}(q^{2}+k^{2}-m^{2}_{K^{*}})^{2}-{1\over3}q^{2}
k^{2}\},\\
&&\frac{d\Gamma^{A}}{dq^{2}}=\frac{G^{2}}{(2\pi)^{3}}\frac{sin^{2}_
{\theta_{C}}}{32m^{3}_{\tau}q^{4}}(m^{2}_{\tau}-q^{2})^{2}
\int^{(\sqrt{q^{2}}-m_{K^{*}})^{2}}_{4m^{2}_{\pi}}dk^{2}
\{{1\over4}(q^{2}+k^{2}-m^{2}_{K^{*}})^{2}-q^{2}k^{2}\}
^{{3\over2}}  \nonumber \\
&&{2\over q^{2}}m^{2}_{\tau}(\frac{N_{C}}
{2\pi^{2}g^{2}})^{2}{1\over\pi}
\frac{\sqrt{k^{2}}\Gamma_{\rho}(k^{2})}{(k^{2}-m^{2}_{\rho})^{2}
+k^{2}\Gamma^{2}_{\rho}(k^{2})},
\end{eqnarray}
where $k^{2}$ is the invariant mass of the two pions and
$\Gamma^{V,A}$ are the decay widths from the vector and axial-vector
currents respectively.

The branching ratio is computed to be $0.24\times10^{-6}$. The
contribution of the axial-vector currents is $17\%$.
The branching ratio of
$\tau\rightarrow K^{*-}\omega\nu$ is determined to be
$0.6\times 10^{-7}$ and the branching ratio of $\tau\rightarrow
\bar{K}^{*0}\rho^{-}\nu$ is $0.48\times 10^{-6}$.

The term(56) is the strong anomaly of the PCAC(\(\Delta s=1\)).
From Eq.(50)
other three terms of the strong anomaly are found
\begin{equation}
-\frac{N_{C}}{2\pi^{2}g^{2}}
\varepsilon^{\mu\nu\alpha
\beta}\{\sqrt{2}
\partial_{\mu}\bar{K}^{*0}_{\nu}\partial_{\alpha}
\rho^{-}_{\beta}+
\partial_{\mu}K^{*-}_{\nu}\partial_{\alpha}
\omega_{\beta}+\sqrt{2}
\partial_{\mu}K^{*-}_{\nu}\partial_{\alpha}
\phi_{\beta}\}.
\end{equation}
Therefore, the PCAC(\(\Delta s=1\)) with strong anomaly is expressed
as
\begin{eqnarray}
\lefteqn{\partial^{\mu}\bar{\psi}\lambda_{+}\gamma_{\mu}\gamma_{5}\psi
=-m^{2}_{K}f_{K}K^{-}}\nonumber \\
&&-\frac{N_{C}}{2\pi^{2}g^{2}}\varepsilon^{\mu\nu\alpha\beta}
\{\partial_{\mu}K^{*-}_{\nu}\partial_{\alpha}\rho^{0}_{\beta}
+\sqrt{2}\partial_{\mu}\bar{K}^{*0}_{\nu}\partial_{\alpha}
\rho^{-}_{\beta}
+\partial_{\mu}K^{*-}_{\nu}\partial_{\alpha}
\omega_{\beta}+\sqrt{2}
\partial_{\mu}K^{*-}_{\nu}\partial_{\alpha}
\phi_{\beta}\}.
\end{eqnarray}
Using the VMD and the substitutions(51),
the electromagnetic anomaly
is added to the Eq.(60)
\begin{eqnarray}
\lefteqn{\partial^{\mu}\bar{\psi}\lambda_{+}\gamma_{\mu}\gamma_{5}\psi
=-m^{2}_{K}f_{K}K^{-}-\frac{e}{2\pi^{2}g}\varepsilon
^{\mu\nu\alpha\beta}\partial_{\mu}K^{*-}_{\nu}\partial_{\alpha}
A_{\beta}}\nonumber \\
&&-\frac{N_{C}}{2\pi^{2}g^{2}}\varepsilon^{\mu\nu\alpha\beta}
\{\partial_{\mu}K^{*-}_{\nu}\partial_{\alpha}\rho^{0}_{\beta}
+\sqrt{2}\partial_{\mu}\bar{K}^{*0}_{\nu}\partial_{\alpha}
\rho^{-}_{\beta}+
\partial_{\mu}K^{*-}_{\nu}\partial_{\alpha}
\omega_{\beta}+\sqrt{2}
\partial_{\mu}K^{*-}_{\nu}\partial_{\alpha}
\phi_{\beta}\}.
\end{eqnarray}
In terms of the soft pion approximation the decay amplitude of
$K^{*-}\rightarrow K^{-}\gamma$ is derived from Eq.(61) in the chiral
limit
\begin{equation}
{\cal M}=-\frac{e}{2\pi^{2}gf_{\pi}}\varepsilon^{\mu\nu\alpha\beta}
p_{\mu}\epsilon_{\nu}(p)k_{\alpha}\epsilon^{*}_{\beta}(k).
\end{equation}
The decay width is found to be
\begin{equation}
\Gamma=\frac{\alpha}{96\pi^{4}g^{2}f^{2}_{\pi}}m^{3}_{K^{*}}(1-{m^{2}
_{K}\over m^{2}_{K^{*}}})^{3}=34.9keV.
\end{equation}
The data[10] is 50.3(1$\pm$0.11)keV. The strange quark mass correction
is responsible for the difference between the theoretical result and
the experiment.

To conclude, the decay $\tau\rightarrow\omega\pi\pi\nu$
is resulted from
the WZW anomaly. Theoretical result of the decay rate agrees
with data. In the chiral limit, the strong anomaly of the PCAC is
found and the strong anomaly originates in the WZW anomaly.
Under the
soft pion approximation the decay rates of $\omega\rightarrow\pi
\gamma$, $\rho\rightarrow\pi\gamma$, and $\omega\rightarrow3\pi$
obtained by using the PCAC with strong anomaly are in good agreement
with data. The strong anomaly of the PCAC leads to the pion dominance
in the decay $\tau\rightarrow\omega\rho\nu$.
The measurements of the decay rate and distribution of
$\tau\rightarrow\omega\rho\nu$ will provide a further evidence on
the strong anomaly of the PCAC.
The strong anomaly of the PCAC(\(\Delta s=1\)) exits too.
It is necessary to emphasize that the Adler-Bell-Jackiw anomaly
and the Adler-Bardeen anomaly of $QCD$ are exact and the strong
anomaly of the PCAC is based on that the meson fields are treated
as effective point fields, like the pion fields in the PCAC and
the meson fields in the WZW anomaly. However, the existence of the
strong anomaly of the PCAC indicates that in the chiral limit,
the quark octet axial-vector currents are not conserved. The
expressions of the strong anomaly presented in this paper, maybe,
are some kind of bosonization of the strong anomaly of the quark
octet axial-vector currents.

The author wishes to thank K.F.Liu for discussion.
This research was partially
supported by DOE Grant No. DE-91ER75661.

{\bf\large Appendix}\\
\begin{equation}
\frac{d\Gamma(\tau\rightarrow\omega\pi\pi\nu)}{dq^{2}dk^{2}
dk^{'2}}=\frac{G^{2}}{256m^{3}_{\tau}}\frac{cos^{2}\theta_{C}}
{(2\pi)^{5}}
{1\over q^{2}}(m^{2}_{\tau}-q^{2})^{2}\{(m^{2}_{\tau}+2q^{2})F
+m^{2}_{\tau}G\},
\end{equation}
where
\begin{eqnarray}
\lefteqn{F=\{{2\over3}{1\over q^{4}}(q\cdot k_{1})^{2}(q^{2}
-q\cdot k_{2})^{2}+{4\over9}{1\over q^{4}}(q\cdot k_{1})^{2}
(q\cdot k_{2})^{2}\}A_{1}}\nonumber \\
&&+\{{2\over3}{1\over q^{4}}(q\cdot k_{2})^{2}(q^{2}
-q\cdot k_{1})^{2}+{4\over9}{1\over q^{4}}(q\cdot k_{2})^{2}
(q\cdot k_{1})^{2}\}A_{2}\nonumber \\
&&+\{{2\over3}{1\over q^{2}}q\cdot k_{1}q\cdot k_{2}q\cdot(k_{1}+
k_{2})-{10\over9}{1\over q^{4}}(q\cdot k_{1})^{2}(q\cdot k_{2})^{2}
\}A_{12}+{2\over9}{1\over q^{4}}(q\cdot k_{1})^{3}(q\cdot k_{2})^{2}
B_{11}\nonumber \\
&&-{2\over9}{1\over q^{4}}(q\cdot k_{1})^{2}(q\cdot k_{2})^{2}
q\cdot(q-k_{2})B_{21}
+{2\over9}{1\over q^{4}}(q\cdot k_{1})^{2}(q\cdot k_{2})^{2}
q\cdot(q-k_{1})B_{12}\nonumber \\
&&-{2\over9}{1\over q^{4}}(q\cdot k_{1})^{2}(q\cdot k_{2})^{3}
B_{22}+{2\over9}{1\over q^{4}}\{(q\cdot k_{1})^{4}(q\cdot k_{2})^{2}
B_{1}+(q\cdot k_{1})^{2}(q\cdot k_{2})^{4}B_{2}\}BW_{a}
\nonumber \\
&&G=[{12\over\pi^{2}g^{2}f^{2}_{\pi}}{c\over g}
(1-{c\over g})]
^{2}{8\over3}
{1\over q^{4}}(q\cdot k_{1})^{2}(q\cdot k_{2})^{2}
\nonumber \\
&&BW_{a}=
\frac{9}{\pi^{4}g^{2}f^{2}_{\pi}}\frac{g^{4}f^{2}_{a}m^{4}_{\rho}
+f^{-2}_{a}q^{2}\Gamma^{2}_{a}(q^{2})}{(q^{2}-m^{2}_{a})^{2}+q^{2}
\Gamma^{2}_{a}(q^{2})},\nonumber \\
&&BW_{\rho}(k^{2})=\frac{1}{(k^{2}-m^{2}_{\rho})^{2}+k^{2}\Gamma
^{2}_{\rho}(k^{2})},\nonumber \\
&&A_{1}=
\{[{2\over f_{\pi}}(1-{2c\over g})+{1\over gf_{a}}(k^{'2}-
m^{2}_{\rho})A(k^{'2})BW_{\rho}(k^{'2})]^{2}+{1\over g^{2}f^{2}_{a}}
A^{2}(k^{'2})k^{'2}\Gamma^{2}_{\rho}(k^{'2})BW_{\rho}(k^{'2})
\},\nonumber \\
&&A_{2}=
\{[{2\over f_{\pi}}(1-{2c\over g})+{1\over gf_{a}}(k^{2}-
m^{2}_{\rho})A(k^{2})BW_{\rho}(k^{2})]^{2}+{1\over g^{2}f^{2}_{a}}
A^{2}(k^{2})k^{2}\Gamma^{2}_{\rho}(k^{2})BW_{\rho}(k^{2})
\},\nonumber \\
&&A_{12}=-2
\{[{2\over f_{\pi}}(1-{2c\over g})+{1\over gf_{a}}(k^{'2}-
m^{2}_{\rho})A(k^{'2})BW_{\rho}(k^{'2})]
[{2\over f_{\pi}}(1-{2c\over g})\nonumber \\
&&+{1\over gf_{a}}(k^{2}-
m^{2}_{\rho})A(k^{2})BW_{\rho}(k^{2})]
+{1\over g^{2}f^{2}_{a}}
A^{2}(k^{'2})A^{2}(k^{2})
\sqrt{k^{2}k^{'2}}
\Gamma_{\rho}(k^{'2})BW_{\rho}(k^{2})
\Gamma_{\rho}(k^{'2})BW_{\rho}(k^{'2})\},\nonumber \\
&&B_{11}=2B\{
[{2\over f_{\pi}}(1-{2c\over g})+{1\over gf_{a}}(k^{'2}-
m^{2}_{\rho})A(k^{'2})BW_{\rho}(k^{'2})]
{1\over gf_{a}}(k^{2}-
m^{2}_{\rho})BW_{\rho}(k^{2})]\nonumber \\
&&+{1\over g^{2}f^{2}_{a}}
A^{2}(k^{'2})
\sqrt{k^{2}k^{'2}}
\Gamma_{\rho}(k^{'2})BW_{\rho}(k^{2})
\Gamma_{\rho}(k^{'2})BW_{\rho}(k^{'2})\},\nonumber \\
&&B_{12}=-2B\{
[{2\over f_{\pi}}(1-{2c\over g})+{1\over gf_{a}}(k^{2}-
m^{2}_{\rho})A(k^{2})BW_{\rho}(k^{2})]
{1\over gf_{a}}(k^{2}-
m^{2}_{\rho})BW_{\rho}(k^{2})]\nonumber \\
&&+{1\over g^{2}f^{2}_{a}}A(k^{2})k^{2}
\Gamma^{2}_{\rho}(k^{2})BW^{2}_{\rho}(k^{2})
\}\nonumber \\
&&B_{21}=2B\{
[{2\over f_{\pi}}(1-{2c\over g})+{1\over gf_{a}}(k^{'2}-
m^{2}_{\rho})A(k^{'2})BW_{\rho}(k^{'2})]
{1\over gf_{a}}(k^{'2}-
m^{2}_{\rho})BW_{\rho}(k^{'2})]\nonumber \\
&&+{1\over g^{2}f^{2}_{a}}A(k^{'2})k^{'2}
\Gamma^{2}_{\rho}(k^{'2})BW^{2}_{\rho}(k^{'2})
\},\nonumber \\
&&B_{22}=-2B\{
[{2\over f_{\pi}}(1-{2c\over g})+{1\over gf_{a}}(k^{2}-
m^{2}_{\rho})A(k^{2})BW_{\rho}(k^{2})]
{1\over gf_{a}}(k^{'2}-
m^{2}_{\rho})BW_{\rho}(k^{'2})]\nonumber \\
&&+{1\over g^{2}f^{2}_{a}}
A^{2}(k^{2})
\sqrt{k^{2}k^{'2}}
\Gamma_{\rho}(k^{'2})BW_{\rho}(k^{2})
\Gamma_{\rho}(k^{'2})BW_{\rho}(k^{'2})\},\nonumber \\
&&B_{1}=B^{2}{1\over g^{2}f^{2}_{a}}BW_{\rho}(k^{2}),
\nonumber \\
&&B_{2}=B^{2}{1\over g^{2}f^{2}_{a}}BW_{\rho}(k^{'2}),
\end{eqnarray}
where \(q=p+k_{1}+k_{2}\),
\(k^{2}=(q-k_{1})^{2}\), \(k^{'2}=(q-k_{2})^{2}\).

\newpage
\begin{center}
Fig.1 Caption
\end{center}
Distribution of the decay rate of $\tau\rightarrow\omega\pi\pi\nu$ vs. the
invariant mass of $\omega\pi\pi$
\begin{center}
Fig.2 Caption
\end{center}
Distribution of the decay rate of $\tau\rightarrow\omega\rho\nu$ vs. the
invariant mass of $\omega\rho$

\end{document}